\documentclass[%
  aps,
  pre,
  superscriptaddress, letterpaper,
  longbibliography,
  showpacs,
  showkeys,
  amsfonts, amssymb, amsmath]{revtex4-1}

\usepackage{xcolor}
\usepackage[applemac]{inputenc}
\usepackage{graphicx,amssymb,amsmath}
\usepackage{epstopdf}							% Converts .eps to
\usepackage{subfig}													% Subfigures
\usepackage{hyperref} % Required for customising links and the PDF
\hypersetup{pdfpagemode={UseOutlines},
bookmarksopen=true,
bookmarksopenlevel=0,
hypertexnames=false,
colorlinks=true, % Set to false to disable coloring links
citecolor=blue, % The color of citations
linkcolor=red, % The color of references to document elements (sections, figures, etc)
urlcolor=black, % The color of hyperlinks (URLs)
pdfstartview={FitV},
unicode,
breaklinks=true,
}

\begin{document}

\title{Confinement effects on gravity-capillary wave turbulence }
\author{Roumaissa Hassaini}
\author{Nicolas Mordant}
\email[]{nicolas.mordant@univ-grenoble-alpes.fr}
\affiliation{Laboratoire des Ecoulements G\'eophysiques et Industriels, Universit\'e Grenoble Alpes, CNRS, Grenoble-INP,  F-38000 Grenoble, France}

\begin{abstract}
The statistical properties of a large number of weakly nonlinear waves can be described in the framework of the Weak Turbulence Theory. The theory is based on the hypothesis of an asymptotically large system. In experiments, the systems have a finite size and the predictions of the theory may not apply because of the presence of discrete modes rather than a continuum of free waves. Our study focusses on the case of waves at the surface of water at scales close to the gravity-capillarity crossover (of order 1~cm). Wave turbulence has peculiar properties in this regime because 1D resonant interactions can occur as shown by Aubourg \& Mordant. Here we investigate the influence of the confinement on the properties of wave turbulence by reducing gradually the size of our wave tank along one of its axis, the size in the other direction being unchanged. We use space-time resolved profilometry to reconstruct the deformed surface of water. We observe an original regime of coexistence of weak wave turbulence along the length of the vessel and discrete turbulence in the confined direction. %We also perform a bicoherence analysis to investigate the nonlinear coupling among the waves. 

\end{abstract}

\maketitle

\section{Introduction}

The dynamics of nonlinear random dispersive waves are involved in many physical systems such as magnetized plasmas \cite{sagdeev19791976}, optics \cite{picozzi2014optical}, thin elastic plates~\cite{During}, capillary waves~\cite{Filonenko} or geophysical flows~\cite{Hasselmann,Galtier}. In the case of nonlinear interactions involving a large number of modes, such systems can evolve into a turbulent state. 
By `turbulent' we mean generically a system involving a large number of degrees of freedom and driven out of equilibrium. In this state, the natural approach is statistical. In the very specific limit of an asymptotically large system and of a weak nonlinearity, the statistical properties of such systems have been studied in the framework of the Weak Turbulence Theory (WTT) since the pioneering work of Zakharov, Benney and Newell in the 60's (see \cite{Nazarenko,newell_wave_2011} for recent reviews). The former hypothesis implies that no discrete modes exist and the latter means that only resonant waves can exchange a significant amount of energy over a finite time. Based on these hypotheses, predictions of the evolution of statistical quantities such as probability distribution or wave spectrum can be derived. For forced and statistically stationary turbulence, the WTT predicts in most cases a direct cascade of energy from the forced scales down to small scales at which dissipation is the most efficient. In many cases, the calculation of the stationary wave spectrum is enabled by the use of the Zakharov transformation~\cite{R1,Nazarenko}.

In this article, we focus on the case of waves at the surface of water. Two asymptotic regimes exist: at large scales the dynamics of the waves is ruled by gravity and at small scales by capillarity. For water, the former is relevant for oceanic waves and the latter is relevant at submillimetric scales. An open question is if and how the energy cascade in the gravity range can connect with a possible energy cascade in the capillary range. Here, we focus on the intermediate scales at the crossover between gravity and capillarity i.e. at centimetric scales. Previous work in this regime by Aubourg \& Mordant~\cite{aubourg_nonlocal_2015,aubourg2016investigation} reported that wave turbulence at these scales was specific in the fact that 3-wave resonant interactions are possible among unidirectional wave packets. This is due to the fact that the dispersion relation of the linear waves is not a pure power law (as opposed to the asymptotic regimes). Indeed the dispersion relation relating the wavenumber $k$ and the angular frequency $\omega$ is
\begin{equation}
\omega = \left(gk + \frac{\gamma}{\rho}k^3\right)^{1/2}
\label{gc_disp}
\end{equation}
where $\gamma$ is the surface tension, $\rho$ is the water density and $g$ is the acceleration of gravity. It evolves from $k^{1/2}$ at low $k$ to $k^{3/2}$ at high $k$ and it is changing its curvature at the capillary-gravity crossover.
Furthermore the unidirectional interactions appear to be dominating the nonlinear activity as compared to interactions between waves propagating at an angle from each other~\cite{aubourg2016investigation}. Thus it opens the possibility of a unidirectional wave turbulence. We stress that such a `1D' wave turbulence involving 3-wave coupling is somewhat exotic. Indeed theoretical works consider most often dispersive waves with dispersion relations that are power laws ($\omega\propto k^\alpha$ with $\alpha > 1$ for resonant 3-wave coupling) that have interesting scaling properties. For such dispersive waves, non trivial 3-wave resonant coupling must involve waves that have distinct propagation directions.

In order to test experimentally if such a `1D' turbulence could exist we reduce the size of our wave tank along one of its axis, keeping the other size unchanged so that to confine progressively the turbulence towards a 2D channel. Thus we are clearly at odds with the hypotheses of the kinetic regime described in the WTT because of the confinement but we remain in the limit of weak nonlinearity. Thus we expect that the nonlinear interaction will remain dominated by resonant or quasi-resonant interactions. 

The direct consequence of finite size is the presence of discrete linear modes that may reduce or even prevent the presence of solutions of the resonance equations such as
\begin{equation}
\mathbf{k_{1}}=\mathbf{k_{2}}+\mathbf{k_{3}}  \quad,\quad \omega_{1}=\omega_{2}+\omega_{3}
\label{system_reson}
\end{equation}
for the case of 3-wave interaction. For instance, Kartashova noted that for pure capillary waves in a square box, no exactly resonant solutions exist~\cite{Kartashova:1994}. For confined gravity waves (4-wave coupling), the number and the size of clusters of interacting waves is strongly depleted~\cite{kartashova2009discrete}. This effect can be overcome if the level of nonlinearity is increased. Indeed due to nonlinearity a spectral widening exists when $T_{NL}$ is finite. The nonlinear spectral width $\Delta \omega^{NL}\propto 1/T_{NL}$ increases with the level of nonlinearity. If the mode separation $\delta \omega$ is small enough then one can have $\Delta \omega^{NL}\gtrsim\delta \omega$ and the energy exchange would not be sensitive anymore to the discreteness of the modes. Thus for a large enough size and a large enough level of nonlinearity one may recover a continuous regime. This transition was analyzed numerically in details by Pan \& Yue~\cite{Pan} in the case of pure capillary waves.

The situation in which the discrete character can not be discarded is called Discrete Wave Turbulence (DWT). Recent works have focused on DWT numerically either in the gravity regime or in the capillary regime~\cite{kartashova2009discrete,l2010discrete,Pan} and experimentally for gravity waves~\cite{denissenko2007gravity,nazarenko2010statistics}. In such a situation, Nazarenko proposed a sandpile scenario: the energy piles up at large scales until the nonlinearity is large enough and triggers an avalanche of energy across scales~\cite{nazarenko2006sandpile}. Then the systems calms down and starts accumulating energy again. If the level of nonlinearity is too weak then a regime of frozen turbulence was  observed numerically for capillary waves~\cite{pushkarev1999kolmogorov}. 
%whereas for gravity waves, even for the smallest nonlinearity, an energy cascade always occurs~\cite{tanaka2004effects}. 
Because our confinement concerns only one axis of the tank, the question is whether our system can still develop a state of continuous interaction in the unconfined direction.

In large enough systems (in all directions) and for weak nonlinearity, the WTT should apply. Let us review shortly observations in the asymptotic regimes : pure gravity waves or pure capillary waves. The WTT predictions for gravity waves are not consistent with the laboratory experiments: The observed spectral exponent of the gravity waves is seen to change strongly with the forcing magnitude and to be close to the WTT predictions only at the highest forcing amplitudes, which appears to be at odds with the hypothesis of weak nonlinearity~\cite{nazarenko2010statistics,falcon2007observation,Aubourg2017}. By contrast, in the ocean, at very large scales, observation of the spectra appear compatible with the theoretical predictions (see for instance recent observations~\cite{Leckler,Lenain}). For capillary waves, experiments seem to follow better the predictions of the WTT especially at strong forcing \cite{wright1997imaging,deike2012decay,Kolmakov,Xia,Punzmann} but much less at weak forcing~\cite{aubourg2016investigation}. 

The reasons invoked to explain the mismatch between theory and experiments are dissipation and finite size. Indeed the theory largely discards the presence of dissipation. The wave equation is usually supposed to be inviscid and dissipation is only used to absorb the energy flux at small scales. In real life, dissipation is usually broadband i.e. not only concentrated at vanishingly small scales. A kinetic regime assumed in the WTT would occur only under the assumption of scale separation: the nonlinear timescale of interaction $T_{NL}$ must be large compared to the wave period $T$ and the dissipation time $T_d$ must be even larger so that:
\begin{equation}
T\ll T_{NL}\ll T_d\, .
\label{sep}
\end{equation}
In water experiments, this may not be the case because of viscous dissipation as shown in \cite{Campagne} or it may require to increase the level of non linearity at the risk of breaking the scale separation $T\ll T_{NL}$. Even if a true scale separation is observed, dissipation may still have an impact on the scaling properties of the wave spectrum as has been shown in \cite{R23,Humbert} for the case of a vibrating elastic plate or in \cite{Deike:2014kt} for gravity-capillarity waves.

We first describe briefly our experiment (part II) and then show the evolution of the Fourier spectrum of the wave field (part III) when the confinement is increased. We then perform a bicoherence analysis to detail the nonlinear coupling among the waves (part IV). Part V is made of a discussion of the results.

\section{Experimental Set-up}
%%%%%%%%%%%%%%%
\begin{table}[!htb]
\center
\begin{tabular}{|l|c|c|c|c|c|c|r|}
  \hline
    case  & $l_0$ & $l_1$ & $l_2$ & $l_3$ & $l_4$\\
  \hline
    width (cm) & 37 & 32.5 & 18.5 & 11 & 7 \\
  \hline
    aspect ratio & 1.5 & 1.8 & 3.1 & 5.2 & 8.1\\
\hline
\end{tabular}
\caption{Values of the confinement that have been used for the variable width tank. Its length is 57~cm and the configurations that we consider in the following (labelled case $l_0$ to $l_4$) correspond to decreasing the width of the from $l_0=37$~cm down to 7~cm. The corresponding aspect ratio is changing from $1.5$ up to $8.1$.}
\label{tableau_confinement}
\end{table}

\begin{figure}[!htb]
\includegraphics[clip,width=14cm]{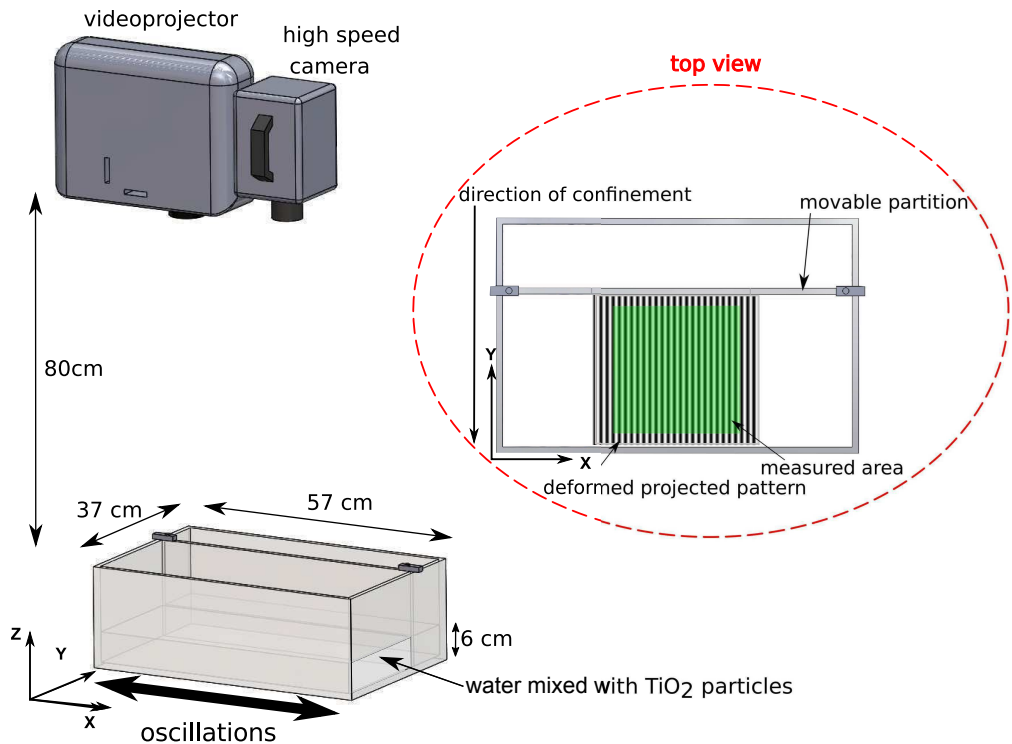} 
\caption{Sketch of the experimental setup. Waves are generated by oscillating horizontally the container. The width of the wave tank can be tuned by a mobile wall so that to modify the horizontal confinement. The wave height is measured by using a Fourier Transform Profilometry technique. See text for details.}
\label{setup}
\end{figure}

We performed experiments in two configurations. The first one is a wave tank whose width can be tuned continuously from 37~cm down to 7 cm (see table \ref{tableau_confinement} for actual values). The length of the tank is 57~cm (along the $x$ axis) so that the aspect ratio of the wave tank increases from $1.5$ up to 8.1. At the lowest confinement it is similar to the one used by Aubourg $\&$ Mordant~\cite{aubourg_nonlocal_2015}. The depth of water at rest is $6$~cm, corresponding to deep water regimes considering  wavelengths of gravity-capillary waves that are less than 10~cm ($kh>1$ where $k$ is the wavenumber and $h$ the depth of water at rest). Great care is taken to prevent surface contamination in order to avoid dissipation by conversion to Marangoni waves~\cite{Przadka}. The waves are generated by exciting continuously horizontally the container (along its long axis $x$) using an oscillating table controlled by a sinusoidal tension modulated around a central frequency equal to  2~Hz with a random modulation in an interval $\pm$ 0.5~Hz around the central frequency. The deformation of the water surface is reconstructed using the Fourier Transform Profilometry \cite{cobelli_global_2009,maurel_experimental_2009} (see fig.~\ref{setup} and \cite{aubourg2016investigation} for more details of the set-up). The water is mixed with Kronos 1001 titanium dioxide particles with a volume fraction of 1$\%$ to improve its optical diffusivity making possible to project on its very surface a sinusoidal grayscale pattern. These particles are chemically neutral and do not alter the surface tension of water as stated by Przadka et al \cite{Przadka}. As the waves propagate the pattern seen by a high speed camera is deformed. The deformation of the pattern can be inverted to provide the elevation field of the waves. In this way we obtain a full space-time resolved measurement of the wave field. The images are recorded at 250 frames/s with a resolution of $1024\times1024$~pixels$^{2}$ covering a $20\times20$~cm$^{2}$ area at the center of the tank in the case $l_0$. When the width of the channel is reduced the resolution is adapted to the width of the channel. The number of pixels along the $y$ axis is reduced accordingly down to a resolution of $360$~pixels keeping 1024 pixels along the $x$ axis.

The second configuration is a high aspect ratio channel of length $1.5$~m and a given width of 10~cm. Its aspect ratio is thus 15. The major interest of this configuration is the much higher length, thus with even less confinement along the $x$ axis. The water depth is the same as well as the measurement method. The images were recorded at the same frequency with a resolution of $1024\times325$~pixels$^{2}$ covering a $31\times10$~cm$^{2}$ area around the center of the channel. 

For all experiments, the average slope of the waves is kept at a value of close to $3\%$ in order to remain in a weakly nonlinear regime. We expect that this regime corresponds to the weak regime identified by \cite{Cobelli}.

\section{Fourier spectra}
%%%%%%%%%%%%

\begin{figure}[htb!]
 (a) \includegraphics[width=10cm]{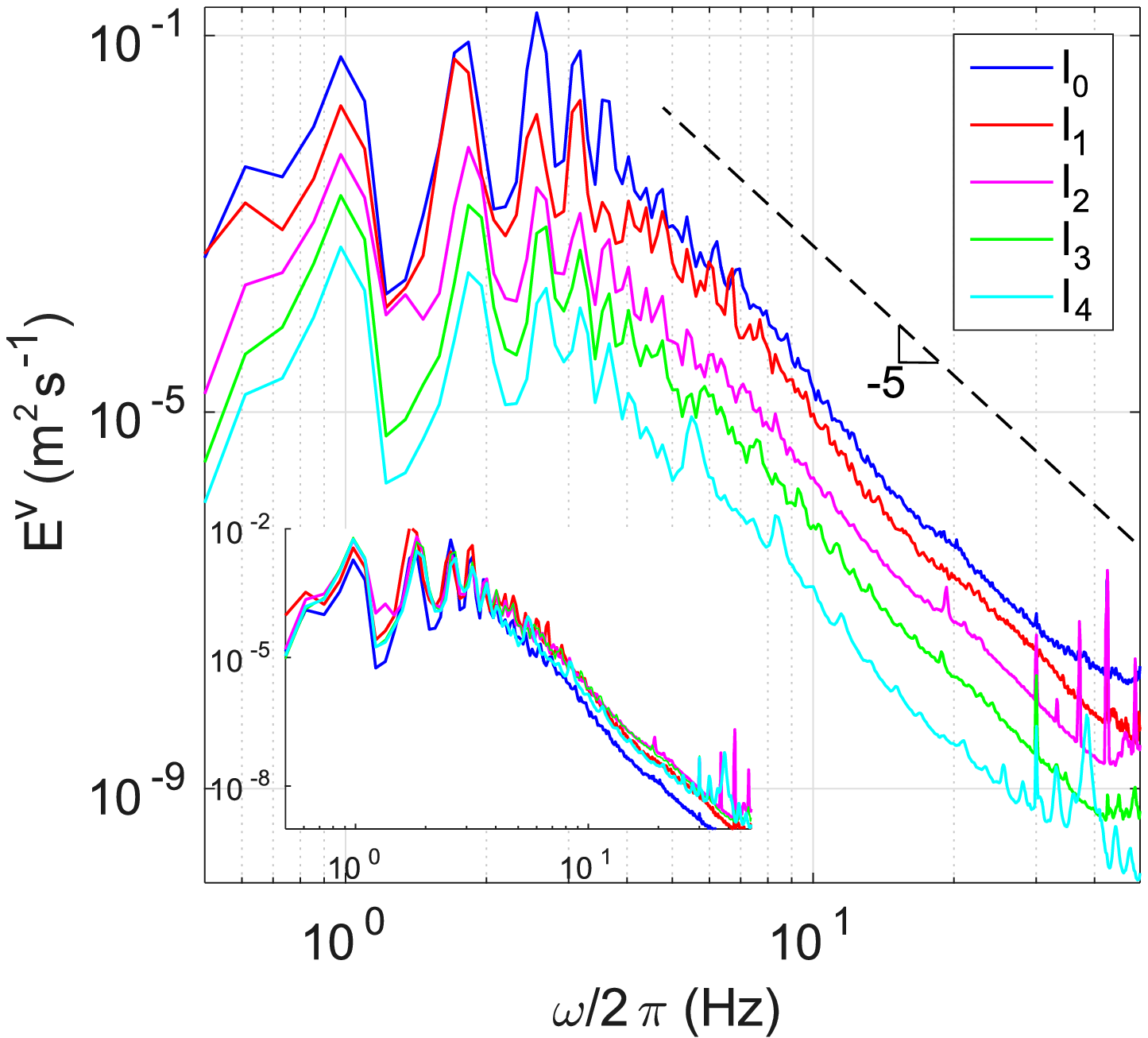}\\
 (b)  \includegraphics[width=9cm]{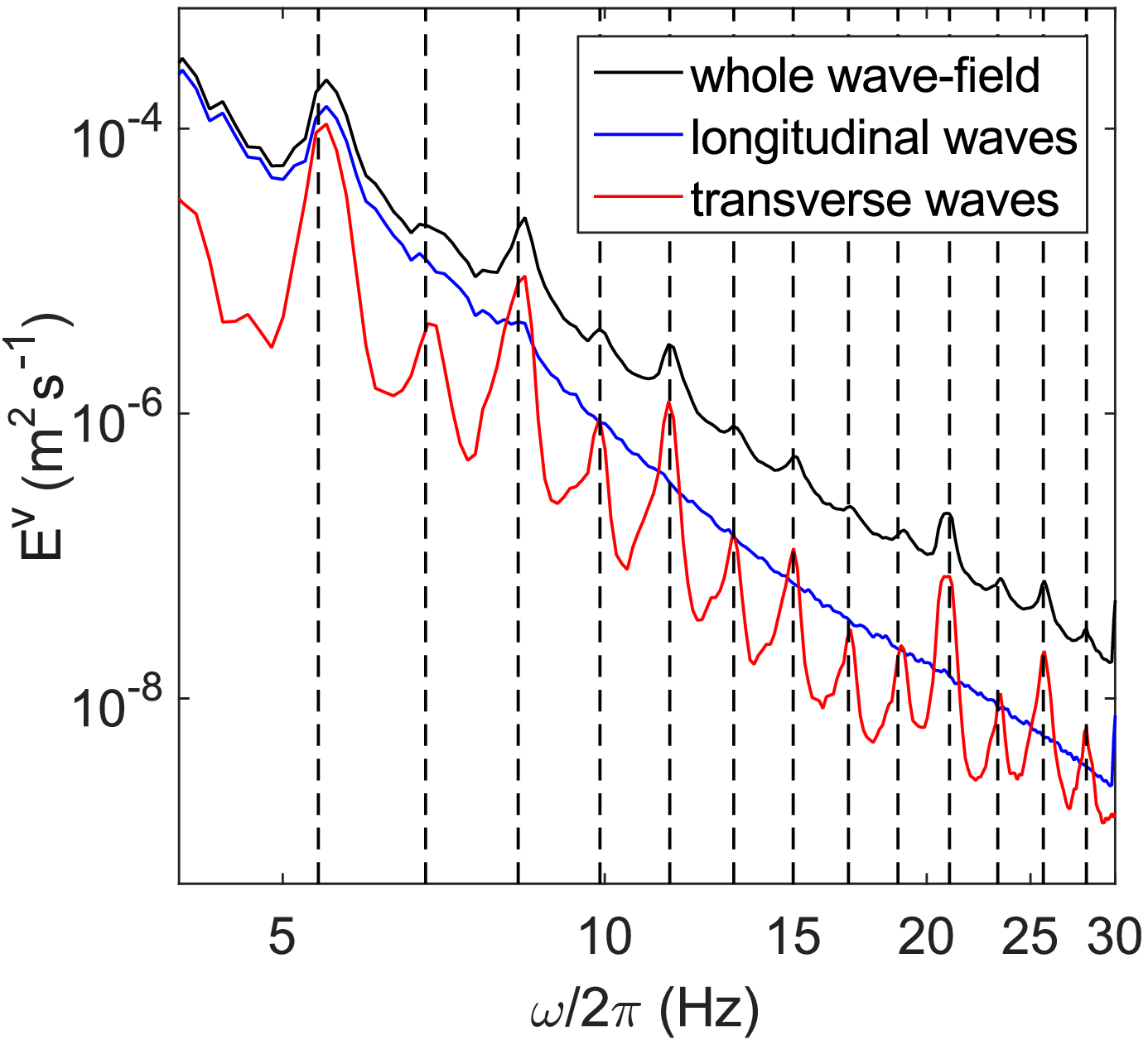}
  \caption{(a) Frequency spectra of the vertical velocity field $E^v(\omega)$ for all confinements.  The spectra have been shifted vertically for clarity in the main figure. The non shifted spectra are shown in the insert. (b) Frequency spectra for the $l_4=7$~cm case. The three curves correspond to the full spectrum (black), the spectrum of the purely longitudinal waves (blue) and the purely transverse waves (red). The vertical dashed lines correspond to the frequency of the transverse modes (table~\ref{eigen}).}
  \label{fsp}
 \end{figure}
 
In a first attempt to identify the effect of confinement, we show in fig.~\ref{fsp}(a) the frequency spectrum of the wave field for all confinements (averaged over all positions). For convenience, we actually compute the spectrum of the vertical velocity $\frac{\partial \eta}{\partial t}$ noted $E^v(\omega)$ ($\eta(x,y,t)$ is the instantaneous height of the free surface). Except for the case of the strongest confinement the decay of the full spectrum is only weakly affected by the confinement. In a frequency band near the forcing (roughly between 1 and 4~Hz) a collection of peaks is visible. It corresponds to a range of development of the cascade. Indeed in the phenomenology of the WTT, the kinetic regime is supposed to occur at frequencies much higher than the forcing ones. This is due to the fact that the phases have to become random due to nonlinear interactions \cite{Nazarenko} whereas the phases in the forcing range are imposed by the shaker. This development of the cascade may occur through the modulational instability of the forcing modes as suggested by Xia~{\it et al.}~\cite{Xia}. At frequencies higher than about 5~Hz the spectrum is smooth for the weakest confined case. This corresponds to an `inertial' range. At frequencies higher than 40~Hz the signal reaches the noise level of the measurement technique. The evolution of the spectrum in the inertial range does not follow a power law decay in contrast with the WTT predictions. It has been discussed in \cite{Campagne} that, for water waves, the viscous dissipation does not allow for a wide enough scale separation between the forcing range and the viscous scales (and the issues is even worse when surface contamination is present). Thus the wave spectrum is affected by dissipation and deviates from the theoretical predictions. This impact of dissipation was also investigated for elastic wave turbulence in a vibrating plate \cite{Humbert,R23}. In our opinion, the power law spectra reported previously by Falcon~{\it et al.} \cite{falcon2007observation} (at least at the strongest forcing for which the scaling is the clearest) correspond most likely to the strong regime identified by Cobelli {\it et al.} \cite{Cobelli} in which coherent structures are present as discussed by Berhanu {\it et al.}~\cite{Berhanu2}. Here we keep the steepness of the waves very low and, as previously reported in \cite{aubourg_nonlocal_2015}, no power law decay is observed. Nevertheless, as investigated previously in \cite{aubourg_nonlocal_2015,aubourg2016investigation}, we trust that we are in a sort of kinetic turbulent regime altered by dissipation. 
 
 A difference is visible between the weakest confined cases and the most confined one ($l_4$):  in the latter some peaks are visible in inertial range emerging from the continuous spectrum (better seen in fig.~\ref{fsp}(b), black curve). In order to have a deeper analysis of this observation we compute the spectrum of the purely longitudinal waves (that propagate along the long axis of the tank, $x$ direction) and the purely transverse ones (along the short axis, $y$ direction). We average the measured wave field $\eta(x,y,t)$ over $y$ to obtain the purely longitudinal part and we average over $x$ for the transverse waves. The corresponding frequency spectra for the case $l_4$ are shown in fig.~\ref{fsp}(b). The longitudinal spectrum (blue curve) remains continuous and monotonous in the inertial range whereas the spectrum of the transverse waves (red curve) shows strongly pronounced peaks that corresponds to the ones visible in the full spectrum (black curve). These peaks correspond to the transverse modes expected from the confinement. Their frequencies can be computed using the discrete wavenumbers $k_y=\dfrac{n\pi}{l_4}$ and the dispersion relation (\ref{gc_disp}):
\begin{equation}
\omega_n = \sqrt{\left(g\frac{n\pi}{l_4}+ \frac{\gamma}{\rho}\left(\frac{n\pi}{l_4}\right)^{3}\right)}
\label{calc_eigen}
\end{equation}
$n$ being an integer related to the order of the mode. The corresponding values are shown in table~\ref{eigen} and shown as vertical dashed lines in fig.~\ref{fsp}(b). Thus we observe an original mixed state of a continuous kinetic-like regime in the longitudinal direction and a discrete regime in the transverse direction.

 \begin{table}[htb!]
\center
  \begin{tabular}{|l|c|c|c|c|c|c|c|c|c|c|c|c|c|c|c|c|c|c|c|c|c|r|}
  \hline
    $n$  & 1 & 2 & 3 & 4 & 5 & 6 & 7 & 8 & 9 & 10 & 11 & 12 & 13 & 14 & 15  \\
  \hline
    $\omega_n/2\pi$ (Hz) & 5.4 & 6.8 & 8.3 & 9.9 & 11.5 & 13.2 & 15.0 & 16.9 & 18.8 & 21 & 23.3 & 25.7 & 28.2 & 307 & 33.3     \\
  \hline
\end{tabular}
\caption{Theoretical frequency of the purely transverse eigen-modes of the tank with the confinement $l_4=7$~cm computed using (\ref{calc_eigen}) and shown in fig.~\ref{fsp}(b).}
\label{eigen}
\end{table}

\begin{figure}[!htb]
(a)  \includegraphics[clip,width=8cm]{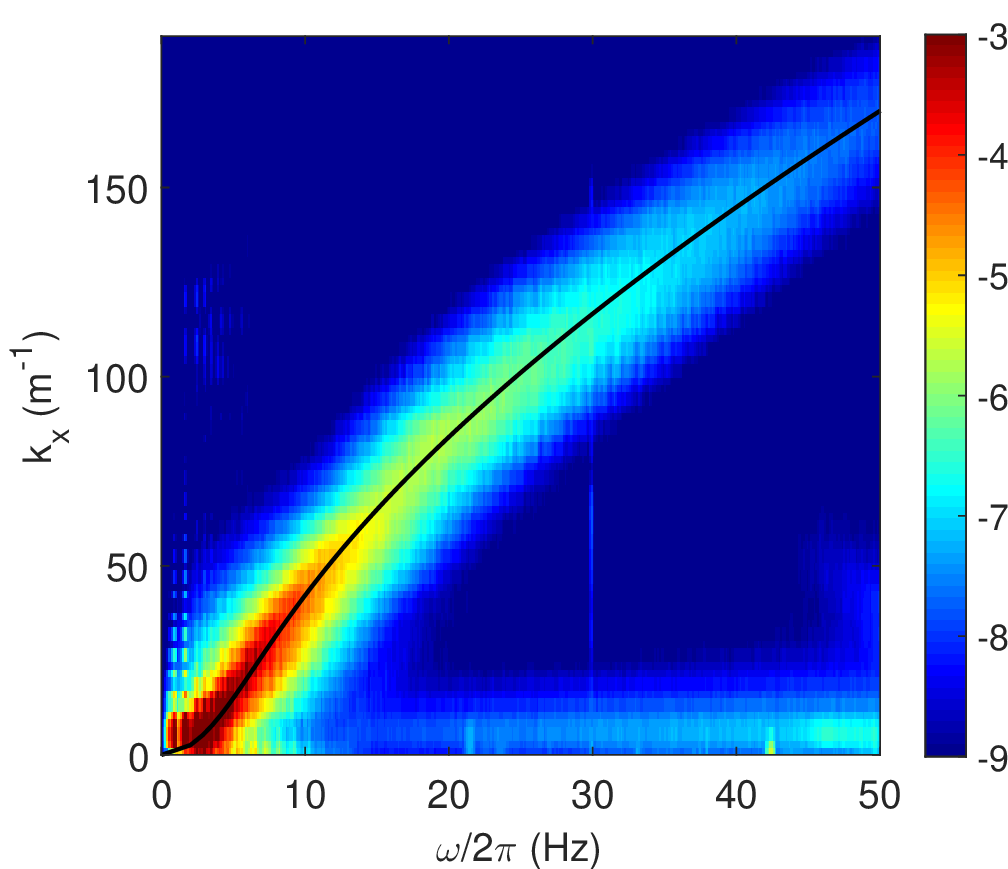} 
(b)  \includegraphics[clip,width=8cm]{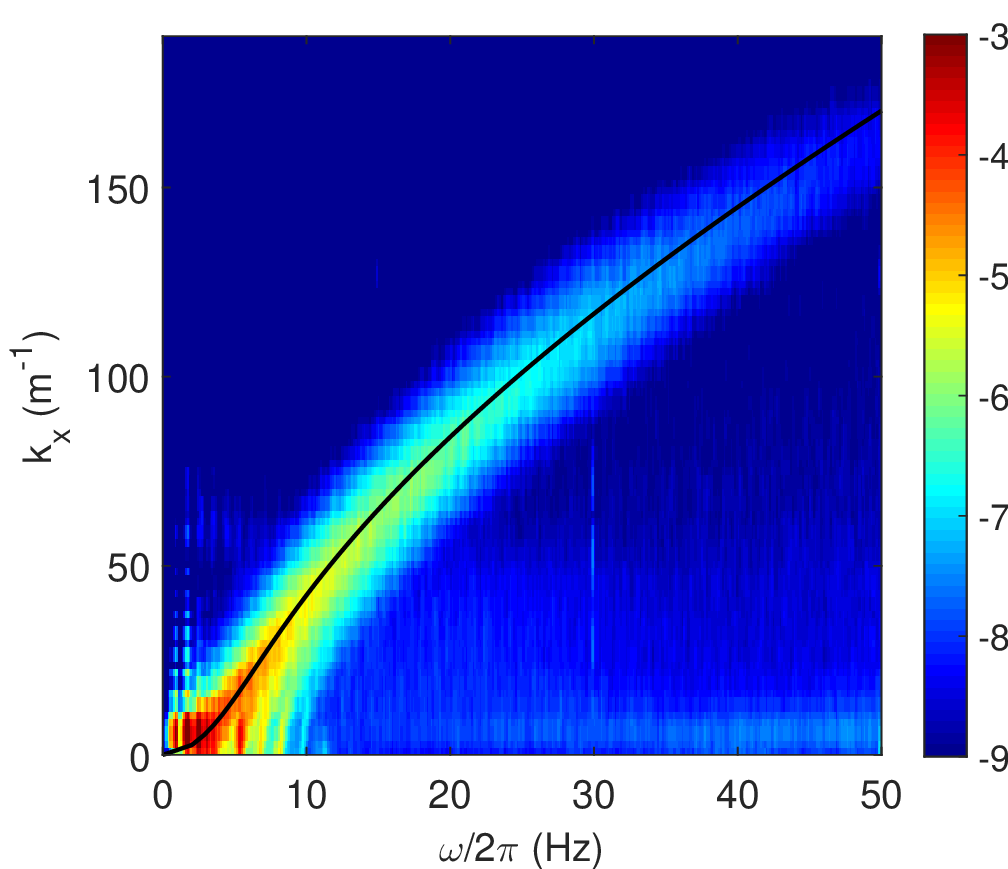}

(c)  \includegraphics[clip,width=8cm]{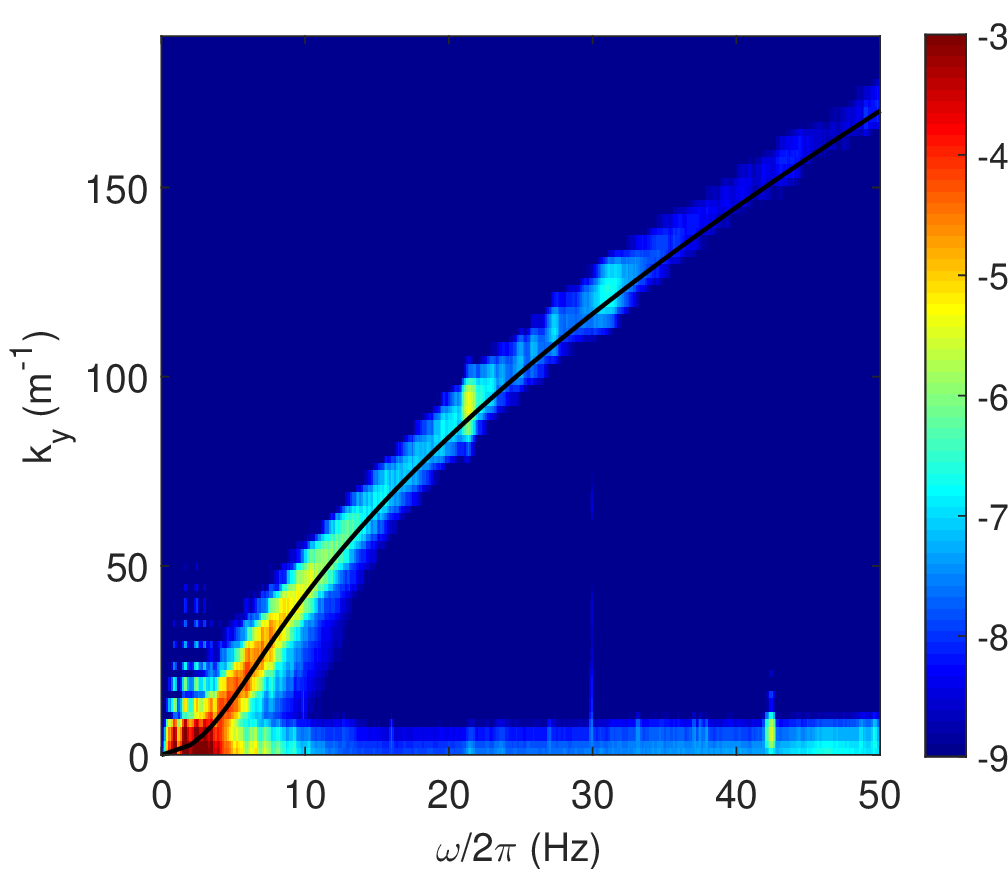}
(d)  \includegraphics[clip,width=8cm]{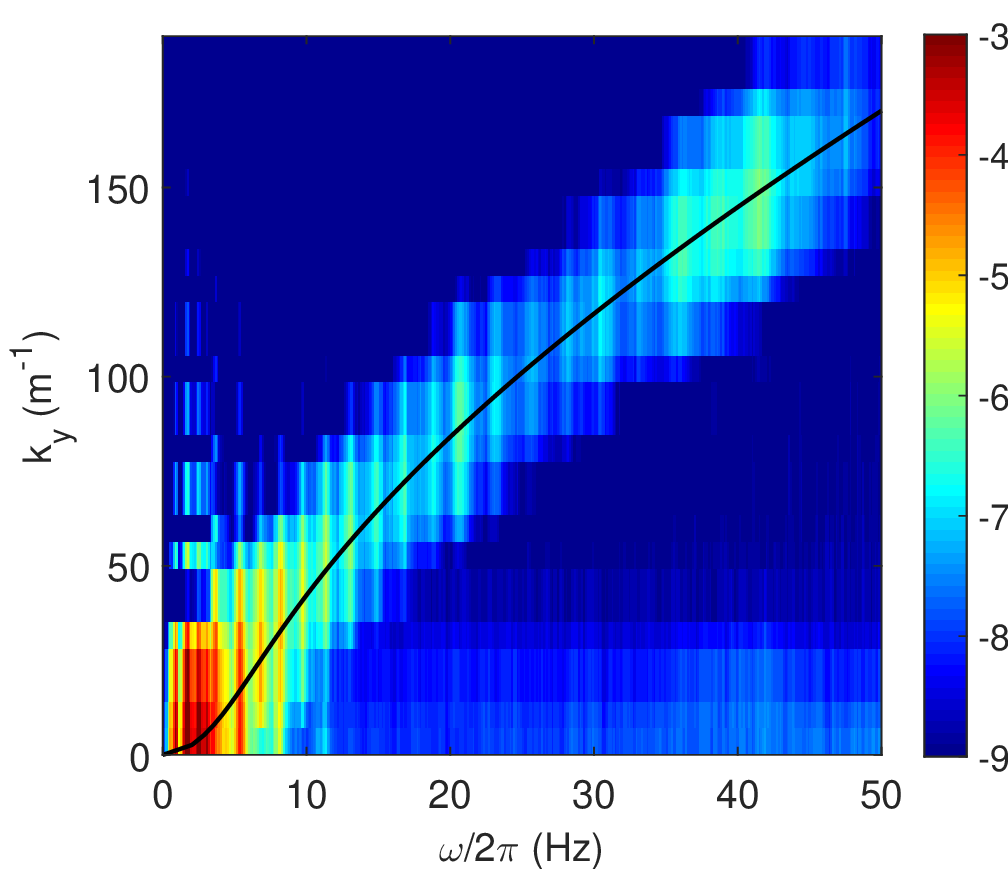}
\caption{Space-time Fourier spectrum of the vertical velocity field of the longitudinal waves $E^v(k_{x},k_{y}=0,\omega)$ ((a) \& (b)) and transverse waves $E^v(k_{x}=0,k_{y},\omega)$ ((c) \& (d)). In (a) \& (c) the width of the tank is equal to $l_1=32.5$~cm. In (b) \& (d)  the width of the tank is equal to $l_4=7$~cm. In all cases, the solid black line is the theoretical deep water linear dispersion relation for gravity-capillary waves in pure water (\ref{gc_disp}). The energy magnitude is displayed in a logarithmic scale.}
\label{spectrakx}
\end{figure}

In order to investigate more precisely the spectral content of the wave field, we compute the full frequency-wavenumber Fourier spectrum noted $E^v(\mathbf k,\omega)$. Similarly to the previous analysis, we distinguish the spectrum $E^v(k_{x},k_{y}=0,\omega)$ of purely longitudinal waves (that propagate along the long axis of the tank, $x$-axis, which is the direction of the forcing as well) from the spectrum of purely transverse waves  $E^v(k_{x}=0,k_{y},\omega)$. The spectra of the longitudinal waves for the unconfined experiment and the $l_4$ case are shown in fig.~\ref{spectrakx}. In the unconfined case (fig.~\ref{spectrakx}(a)\& (c)) the spectra of transverse and longitudinal waves are quite similar: the energy is continuously spread along the dispersion relation of deep water waves (\ref{gc_disp}) with $\gamma= 72$~mN/m the surface tension of pure water, $\rho=1000$~kg/m$^{3}$ and $g=9.81$~m/s$^2$. The fact that the forcing is along the $x$ axis shows up in the fact that the energy of transverse waves is smaller but a directional redistribution of energy is operating as observed in our previous work \cite{aubourg_nonlocal_2015,aubourg2016investigation,hassaini2017transition}. These observations are qualitatively compatible with the direct cascade of energy predicted by the theory~\cite{Filonenko} and these spectra are consistent with observations reported in \cite{Berhanu,Cobelli} for instance. By contrast, the longitudinal and transverse spectra observed in the confined case (fig.~\ref{spectrakx}(b)\& (d)) are very different. The spectrum of longitudinal waves is almost unchanged but the transverse spectrum is now made of well separated peaks corresponding to the modes of the container.  
It should be noted as well that a broadening of the spectrum is observed in this case of strong confinement due to a reduced resolution in $k_y$ related to a smaller width of the images that are bounded by the walls in the $y$ direction. The resolution in the $k_x$ direction is not altered as the size of the image is unchanged along the $x$-axis.

\begin{figure}[!htb] 
(a)\includegraphics[clip,width=11.5cm]{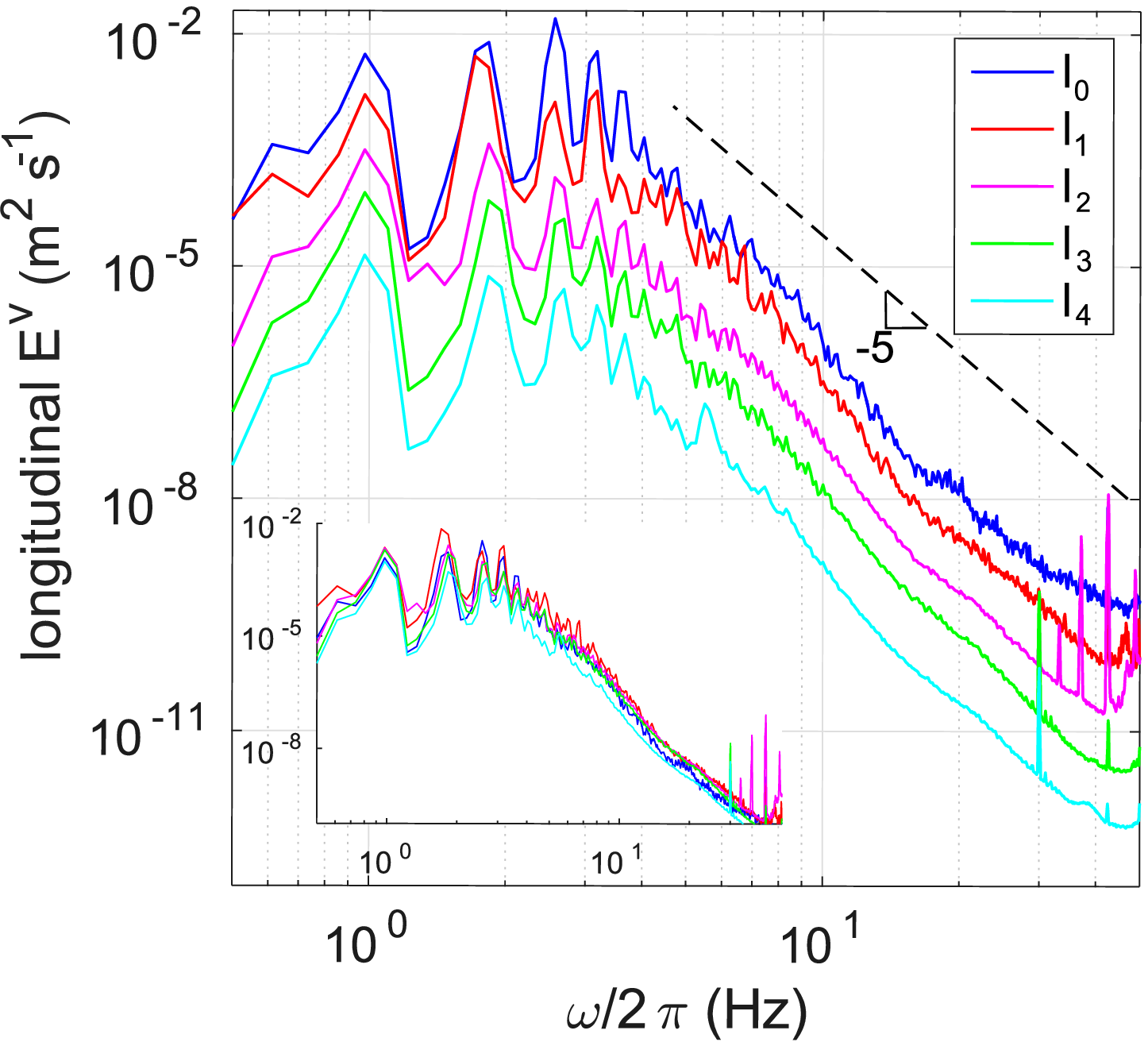}\\
(b)\includegraphics[clip,width=11.5cm]{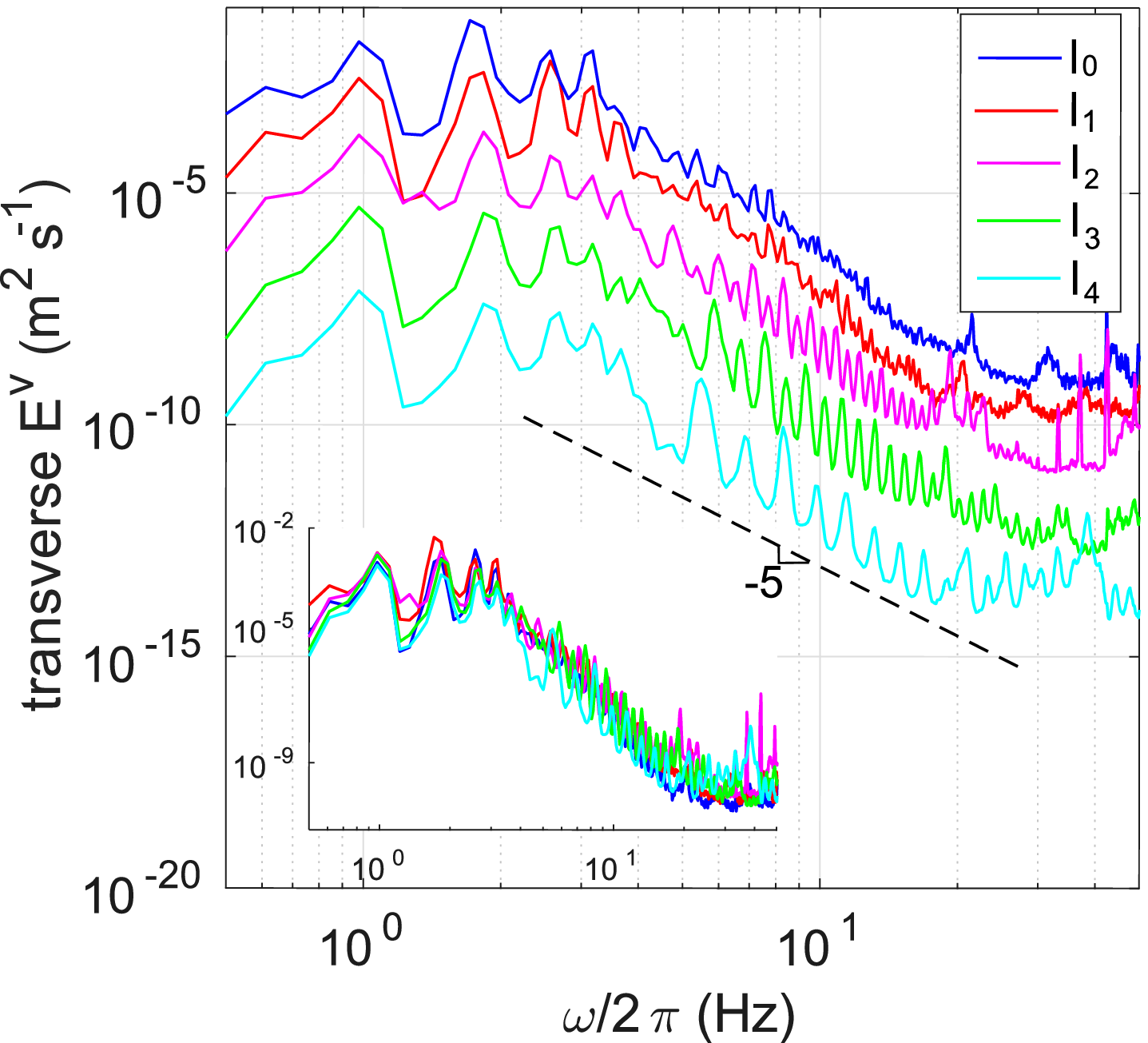}
\caption{ Frequency spectra of the longitudinal (a) and transverse (b) waves for all confinements. The width of the vessel is indicated in the legend. The spectra have been shifted vertically for a better visualization, with confinement increasing from the top to the bottom. The inserts show the unshifted curves. The dashed line is a $f^{-5}$ as an eye guide.}
\label{pente}
\end{figure}
To observe more progressively the emergence of the discretization of the spectra, we display in fig.~\ref{pente} the frequency spectra $E^v(\omega)$ of longitudinal and transverse waves for all values $l_i$ of the width of the channel . 
For longitudinal waves the frequency spectra are continuous and similar in all cases, as previously shown in fig.~\ref{spectrakx}. Only the most confined case shows a slightly distinct decay of the energy but even for this configuration, the spectrum remains continuous. 

For transverse waves (fig.~\ref{pente}(b)) we observe a transition from a continuous spectrum (for $l_0$ and $l_1$) to a discrete spectrum for the more confined cases. The peaks of the spectrum become more and more separated as the confinement increases as expected from (\ref{calc_eigen}). The spectrum becomes discrete when the spectral width between peaks becomes larger than the width of a single peak. This observation is qualitatively consistent with the predictions of~\cite{kartashova2009discrete,nazarenko2006sandpile} although the exact qualification of our state between `discrete', `frozen' or `sandpile' is unclear thus we use the word `discrete' as generic term. In particular the interactions of transverse waves and oblique or longitudinal ones makes the phenomenology more complex.

In the framework of the Weak Turbulence Theory, for simple dispersion relations behaving as $\omega\propto k^\beta$, the predicted energy spectrum E($k$) can often be written as:
\begin{equation}
 E(k) =C P^{1/(N-1)}k^{-\alpha}
\label{spectrum_k}
\end{equation}
where $ k = \| \mathbf{k} \|$ is the wave number, $P$ the energy flux, $C$ a dimensional constant and $\alpha$ the spectral exponent \cite{Nazarenko}. $N$ is the number of waves taking part in the resonances: Indeed one of the hypotheses of WTT is that energy is transmitted among resonant waves. At the lowest order, it involves 3 waves that  have then to satisfy the resonance conditions (\ref{system_reson}).

In the case of gravity waves the resonance conditions do not admit solutions for 3 waves because of the negative curvature of the dispersion relation in deep water  $\omega = \sqrt{gk}$. One needs to consider the next order which involves 4 waves. For the capillary domain the dispersion relation for infinite depth is $\omega = (\frac{\gamma}{\rho})^{\frac{1}{2}}\,k^{\frac{3}{2}}$ and 3-wave resonances exist. The predicted spectra for water waves are thus expected to be $E(\omega) \propto P^{1/3}g\omega^{-4}$ for gravity waves and $E(\omega) \propto P^{1/2}\left(\frac{\gamma}{\rho}\right)^{1/6}\omega^{-17/6}$ for capillary waves~\cite{Filonenko}. 
In our experiment, we do not observe power laws but a decay which is faster than $f^{-5}$ (fig.~\ref{pente}) as is often observed in experiments (due to finite size or dissipation as discussed above)\cite{nazarenko2010statistics,falcon2007observation,Deike,Campagne,Aubourg2017} and the steepness of the spectrum is not depending on the confinement.
%Having for our experiments a very weak nonlinearity around $3\%$ explains why we observe steeper energy spectra than expected by the theory. Note that dissipation is also well known to cause steeper spectra than predicted by the theory [REFs].  

\begin{figure}[!htb]
(a)  \includegraphics[clip,width=7cm]{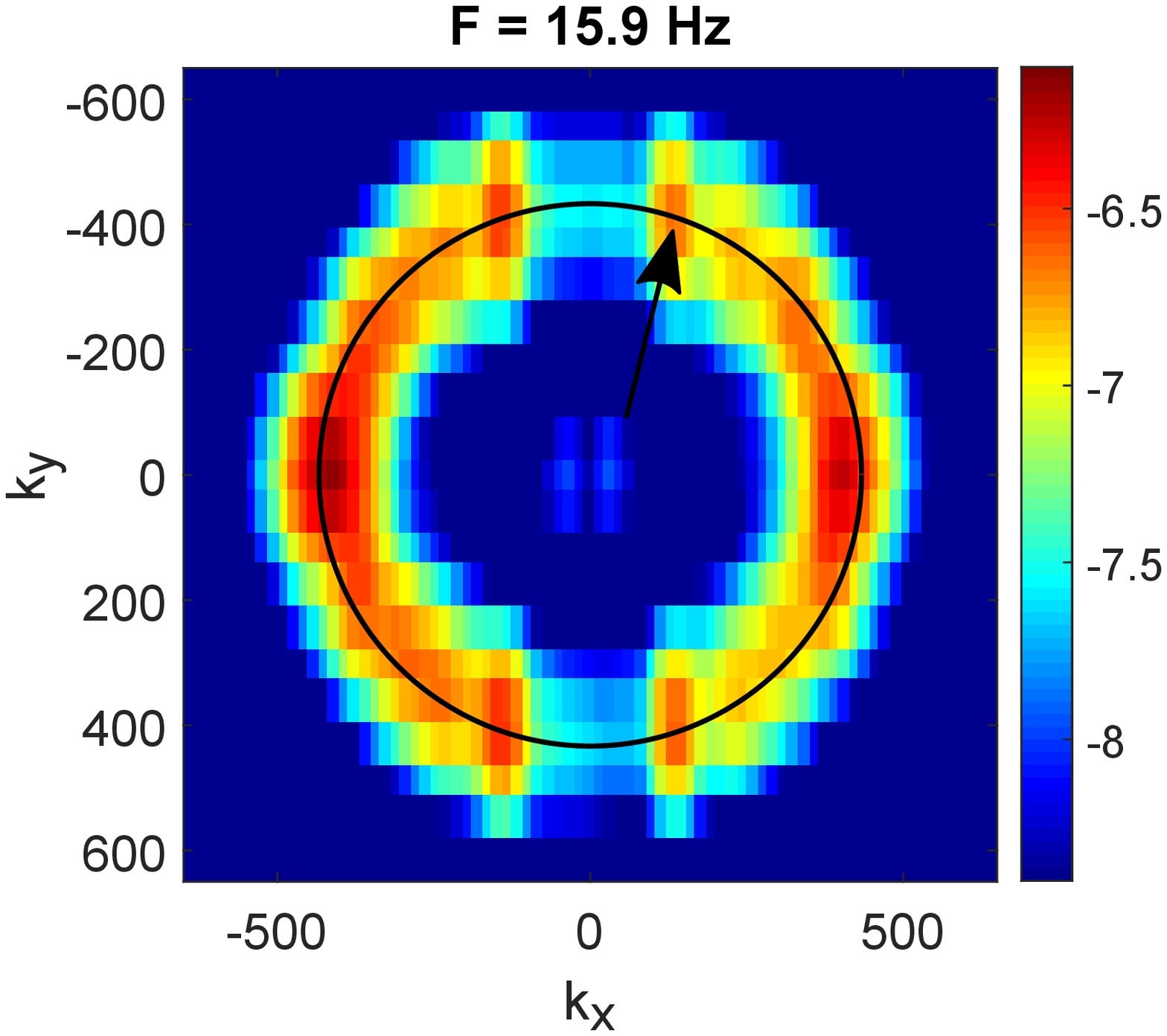}
(b)  \includegraphics[clip,width=7cm]{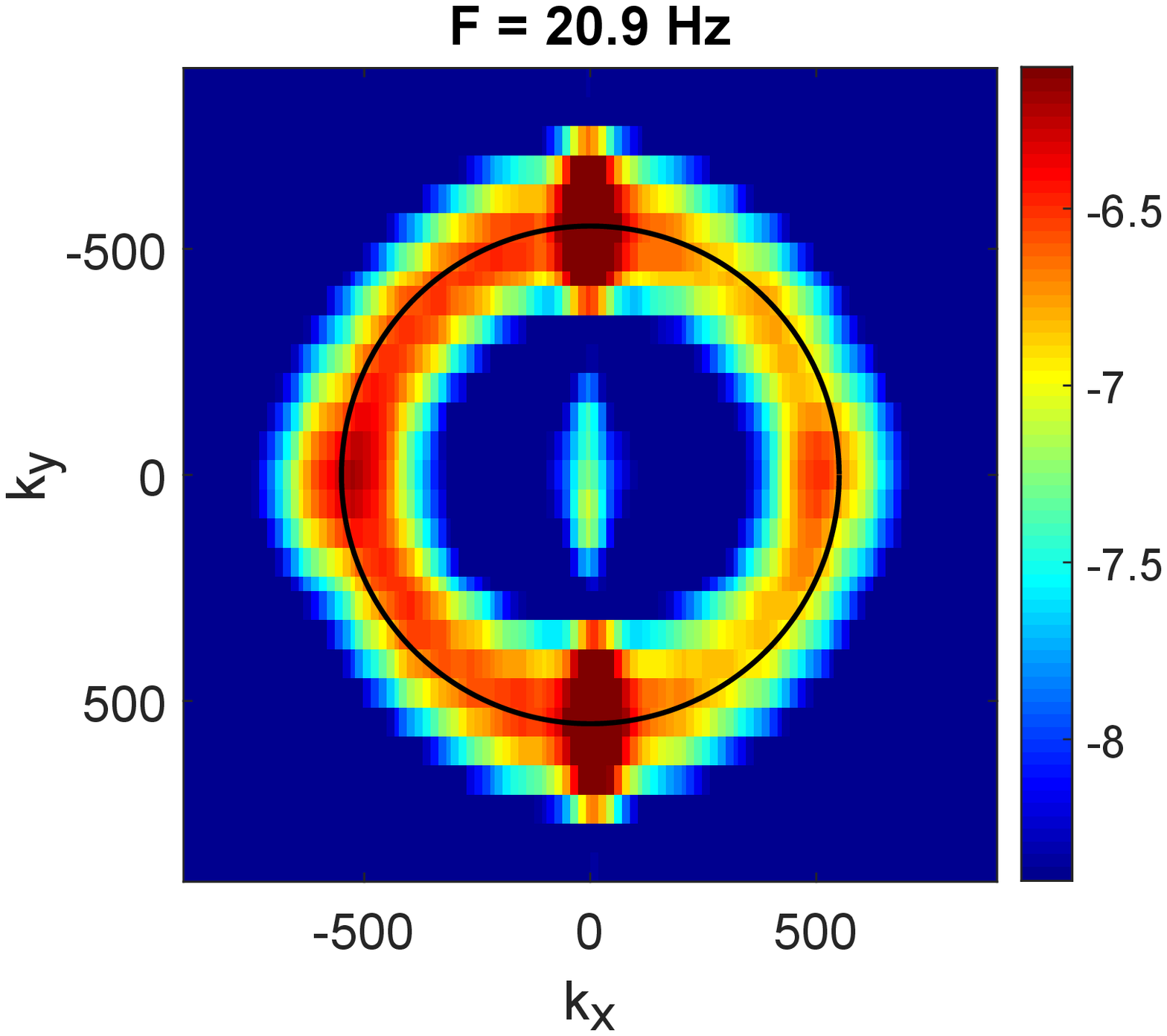}
\caption{(a) cut of the spectrum for the full wave field for the maximum confinement at a frequency which lies in between two successive transverse eigenmodes of the vessel $E^v(k_{x},k_{y},\frac{\omega}{2\pi}=15.9Hz)$. (b) cut of the spectrum for the full wave field for the maximum confinement at a frequency which corresponds to a transverse eigenmode of the vessel $E^v(k_{x},k_{y},\frac{\omega}{2\pi}=20.9Hz)$. In both (a) and (b), the solid black circle is the theoretical deep water linear dispersion relation for gravity-capillary waves in pure water. The energy magnitude is in a logarithmic scale.}
\label{isotropy}
\end{figure}
Let us focus now on the directional properties of the wave field. We show in fig.~\ref{isotropy} a cut of the spectrum $E^v(\mathbf k,\omega)$ for two given values of the frequency in the case of strongest confinement. In (a) the frequency has been chosen in between the frequencies corresponding to two successive transverse modes. In (b) the given frequency is that of a transverse mode. As expected from fig.~\ref{spectrakx} the energy is located on a circle that corresponds to the linear dispersion relation (black circle). 
For both (a) and (b) the energy is quite isotropically distributed except near the purely transverse waves corresponding to small values of $k_x$. In (a) a strong dip is observed at $k_x=0$ related to the fact that there is no purely transverse mode at that frequency. A peak of energy is observed at a position slightly turned from the $k_y$ axis that corresponds to a discrete oblique mode (black arrow). Modes that would be more oblique are not visible as the spectrum is then continuous for directions closer to the $k_x$ axis. In (b) a strong peak is observed at $k_x=0$ corresponding to the transverse mode that resonates accross the channel. 

\begin{figure}[!htb] 
\includegraphics[clip,width=10cm]{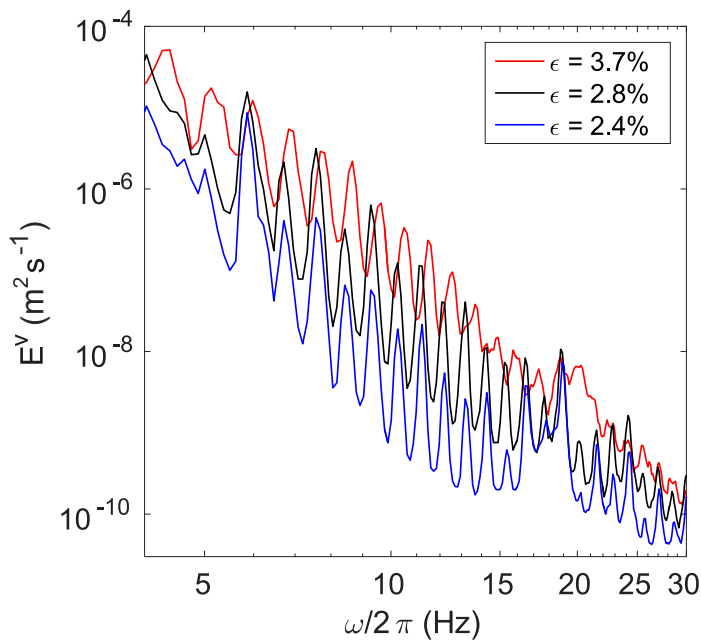}
\caption{Frequency spectra of the transverse waves in the case $l_{3}$ with nonlinearity (average slope) $\epsilon$ increasing from the bottom to the top.}
\label{spectre_nonl}
\end{figure}
Figure~\ref{spectre_nonl} shows the effect of the amplitude of the forcing on the transverse spectrum of the waves in case $l_3$. Increasing the nonlinearity induces two effects. First the height of the peaks is lowest (or the trough between adjacent peaks is less deep) at strongest forcing due to a widening of the peaks. Second, the frequency of the peaks is slightly modified most likely due to non linear corrections on the dispersion relation as that reported by Berhanu {\it et al.}~\cite{Berhanu}. A similar feature was also reported for wave turbulence in an elastic plate~\cite{R21}. The widening of the peak is due to a decrease of the non-linear timescale. For close enough peaks or strong enough a nonlinearity one expects that the peaks will merge and the spectrum will evolve into a continuous spectrum (as seen for capillary waves by Pan \& Yue \cite{Pan} and for elastic waves by Mordant~\cite{R21}). This may not happen if it requires a too strong level of nonlinearity so that other phenomena such as drop ejection may occur that would change significantly the underlying physics.

Thanks to the Fourier spectra, we were able to identify that when the confinement increases in one direction, we still have a weak turbulent regime for the waves that propagate in the unconfined direction with a typical continuous aspect whereas the energy spectra of the waves propagating in the confined direction are distributed only on the discrete transverse eigenmodes of the vessel. A coexistence between weak turbulence and discrete turbulence is clearly observed. 

We will focus in the next part on the nonlinear interactions between resonant waves.

\section{3-Wave interaction} 
%%%%%%%%%%%%%%

\begin{figure}[!htb]
\includegraphics[clip,width=7.5cm]{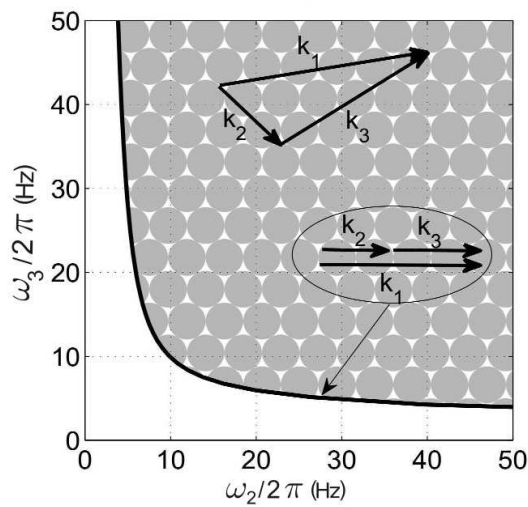}
\caption{Domain of existence of resonant waves (\ref{system_reson}) in the frequency domain for gravity-capillary waves. The solid black line is the border of this domain and corresponds to interactions between three copropagating waves. The gray pattern area above the solid black line represents solutions where the directions of the three waves are no longer the same. In the white area below the solid black line, no resonant solutions exist.}
\label{geom_3W}
\end{figure}
As briefly discussed in the previous paragraph, one of the outcomes of WTT is that energy is transmitted among resonant waves. At the lowest order, for pure capillary waves~\cite{Filonenko}, it involves 3 waves that have then to satisfy the resonance conditions (\ref{system_reson}). 
It is possible to find geometrical solutions to this system using the linear dispersion relation of gravity-capillary waves. Fig.~\ref{geom_3W} shows the domain of existence of resonant waves in $(\omega_2,\omega_3)$ variables. For too small values of either frequency, no 3-wave resonant solution can exist. This is related to the fact that no resonance can exist among 3 purely gravity waves due to the curvature of the dispersion relation. For large enough values of $\omega_2$ and $\omega_3$ solutions exists as is the case for purely capillary waves. A border exists (black line) above which resonant solutions can exist that correspond to 3 wave propagating in distinct directions~\cite{theseAubourg2016,aubourg_nonlocal_2015}. At the border, waves propagate in the same direction.

%In the case of Discrete Turbulence Theory the energy is transmitted among quasi-resonant waves. Indeed since the discrete modes are sparsely separated in the k-space an energy widening around them is needed in order to trigger resonances. The conditions of resonances for a triad need then to be rewritten as $\mathbf{k_{1}}=\mathbf{k_{2}}+\mathbf{k_{3}}\pm \delta\mathbf{k}$  and  $\omega_{1}=\omega_{2}+\omega_{3}$. $\pm \delta\mathbf{k}$ being the widening of the experimental spatio-temporal spectrum in the k-space.

To investigate the 3-wave coupling in our data, we compute third order correlations of the velocity field. From $v(x,y,t)$, we compute the Fourier transform in time over 4~s time windows so that to obtain $v(x,y,\omega)$. 3-wave bispectra or rather bicoherences (i.e. dimensionless bispectra) are then computed as
\begin{equation}
B(\omega_{2},\omega_{3}) = \frac{|\left\langle\left\langle v^{\ast}(x,y,\omega_{2} + \omega_{3})v(x,y,\omega_{2})v(x,y,\omega_{3})\right\rangle\right\rangle|}{[E^{v}(\omega_{2} + \omega_{3})E^{v}(\omega_{2})E^{v}(\omega_{3})]^{\frac{1}{2}}}
\label{C3}
\end{equation}
where $^{\ast}$ stands for complex conjugation and the average $\langle\langle...\rangle\rangle$ stands for an average over successive time windows and a space average over (x,y) positions on the image. $E^{v}(\omega) = \langle\langle|v(x,y,\omega)|^{2}\rangle\rangle$ is the frequency spectrum. 

\begin{figure}[!htb]
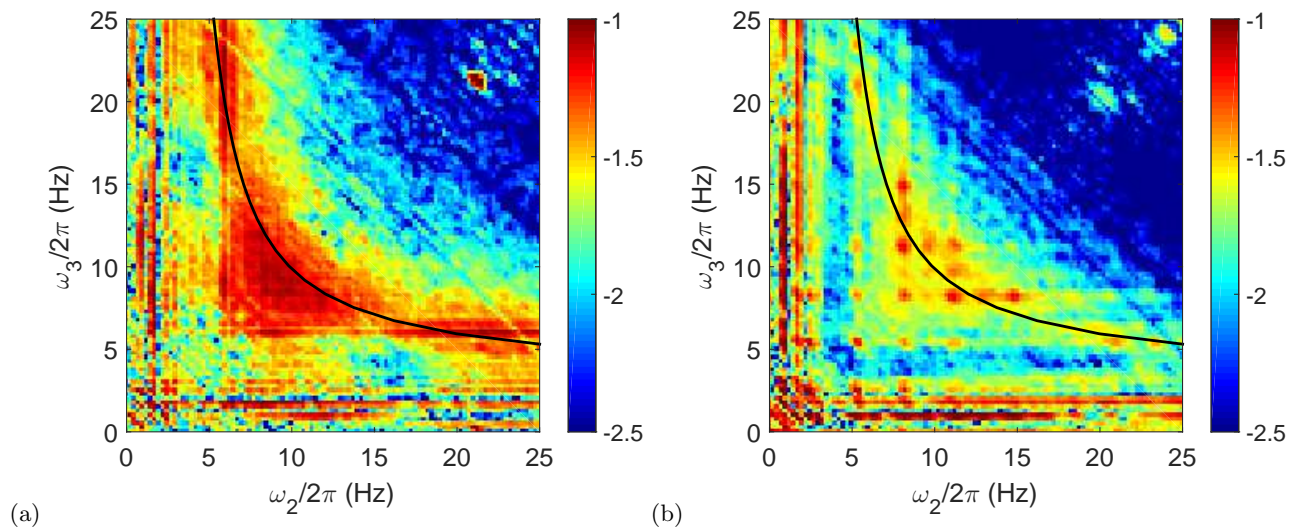

(a)\includegraphics[clip,width=8cm]{bicoherence_l1.eps}
(b)\includegraphics[clip,width=8cm]{bicoherence_l4.eps}
\caption{Bicoherence (\ref{C3}) of the velocity field for (a) the experiment with the weakest confinement $l_1=32.5$~cm and (b) for the experiment with the strongest confinement $l_4=7$~cm. The black line corresponds to the 1D 3-wave resonant solutions.}
\label{B2brut}
\end{figure}
Fig.~\ref{B2brut}(a) displays the bicoherence map of the full velocity field  for experiments for the weakest confinement (case $l_0$). We can see a line of strong correlation along the line of unidirectional coupling (black line) as well as some significant correlations in the region where supposedly no correlations are possible (below the black line). These features have been reported and discussed in a previous article by Aubourg \& Mordant~\cite{aubourg_nonlocal_2015}. In particular the correlations observed in the ``forbidden region'' are due to quasi-resonnances allowed by the finite level of nonlinearity. The bicoherence measured for the strongest confinement $l_4=7$~cm is shown in fig.~\ref{B2brut}(b). It shows a quite different organization of the signal with a dotted pattern rather than a continuous line as observed in (a). 

\begin{figure}[!htb]
(a)\includegraphics[clip,width=10.5cm]{bicoherence_l4_x.eps}
%(b)\includegraphics[clip,width=8cm]{bicoherence_l4_y.eps}
(b)\includegraphics[clip,width=11cm]{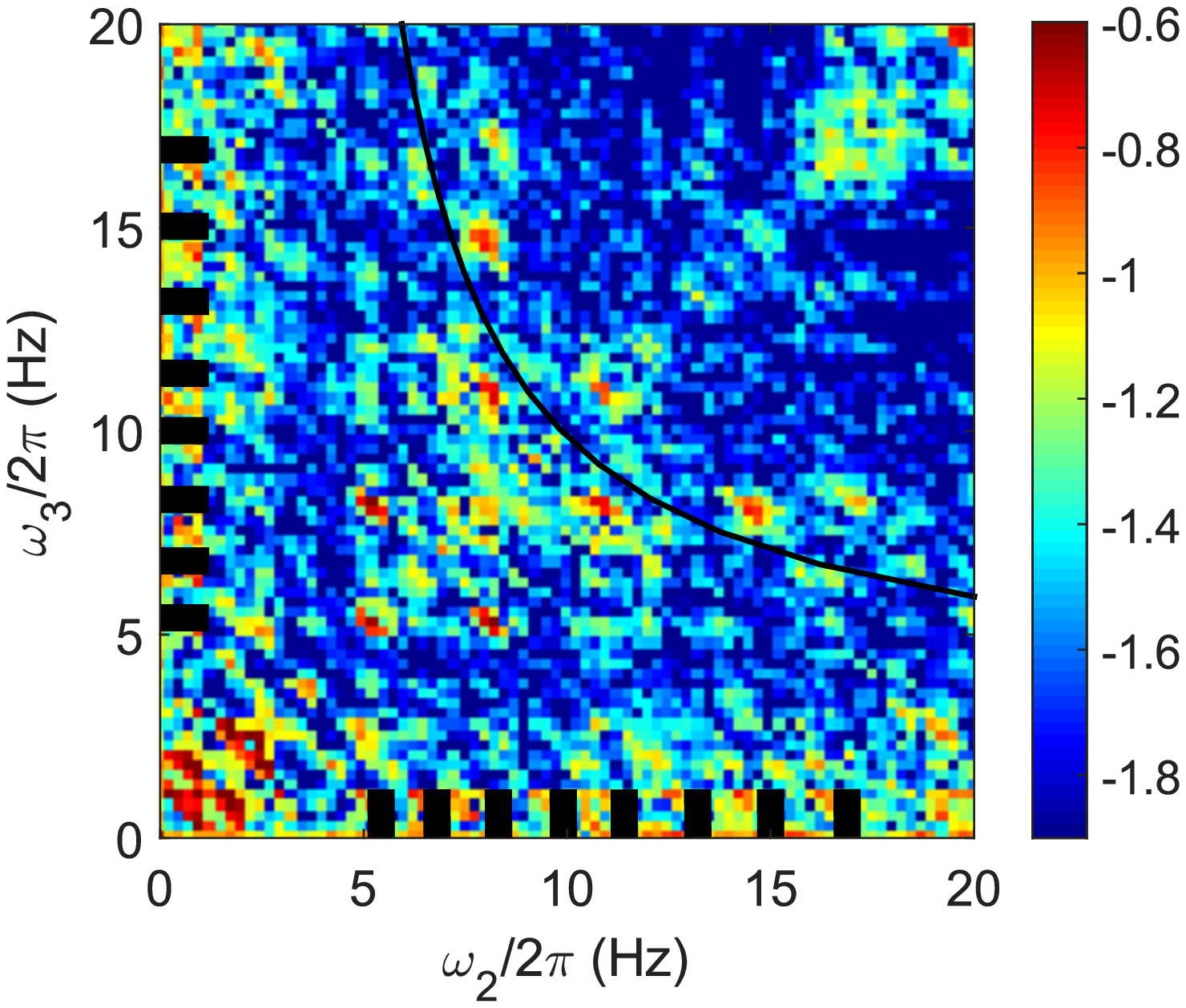}
\caption{Bicoherence of the longitudinal waves (a) and transverse waves (b) for the experiment with the most important confinement corresponding to a width of the vessel equal to 7~cm. The black line corresponds to the 1D solutions. The frequencies associated with the discrete transverse modes appear as black rectangle in (b).}
\label{B2discrete}
\end{figure}
To better understand this observation we compute separately the bicoherence of the velocity of longitudinal  and transverse waves. Both cases are shown in fig~\ref{B2discrete}. 
 As it was observed for the spatiotemporal spectra where the longitudinal spectrum was not affected by the confinement, the same observation is seen for the bicoherence of the longitudinal waves which looks similar to the bicoherence of the unconfined case. Note that the statistical convergence of the bicoherence is lower than that of the full field as less wave configurations are averaged.
By contrast, the bicoherence of the transverse waves shows only interactions among the scattered transverse eigenmodes of the vessel. For some unclear reasons, only even order modes appear in the bicoherence.

\begin{figure}[!htb]
(a)\includegraphics[clip,width=8cm]{bicoherence_canal.eps}

(b)\includegraphics[clip,width=8cm]{bicoherence_canal_x.eps}
(c)\includegraphics[clip,width=8cm]{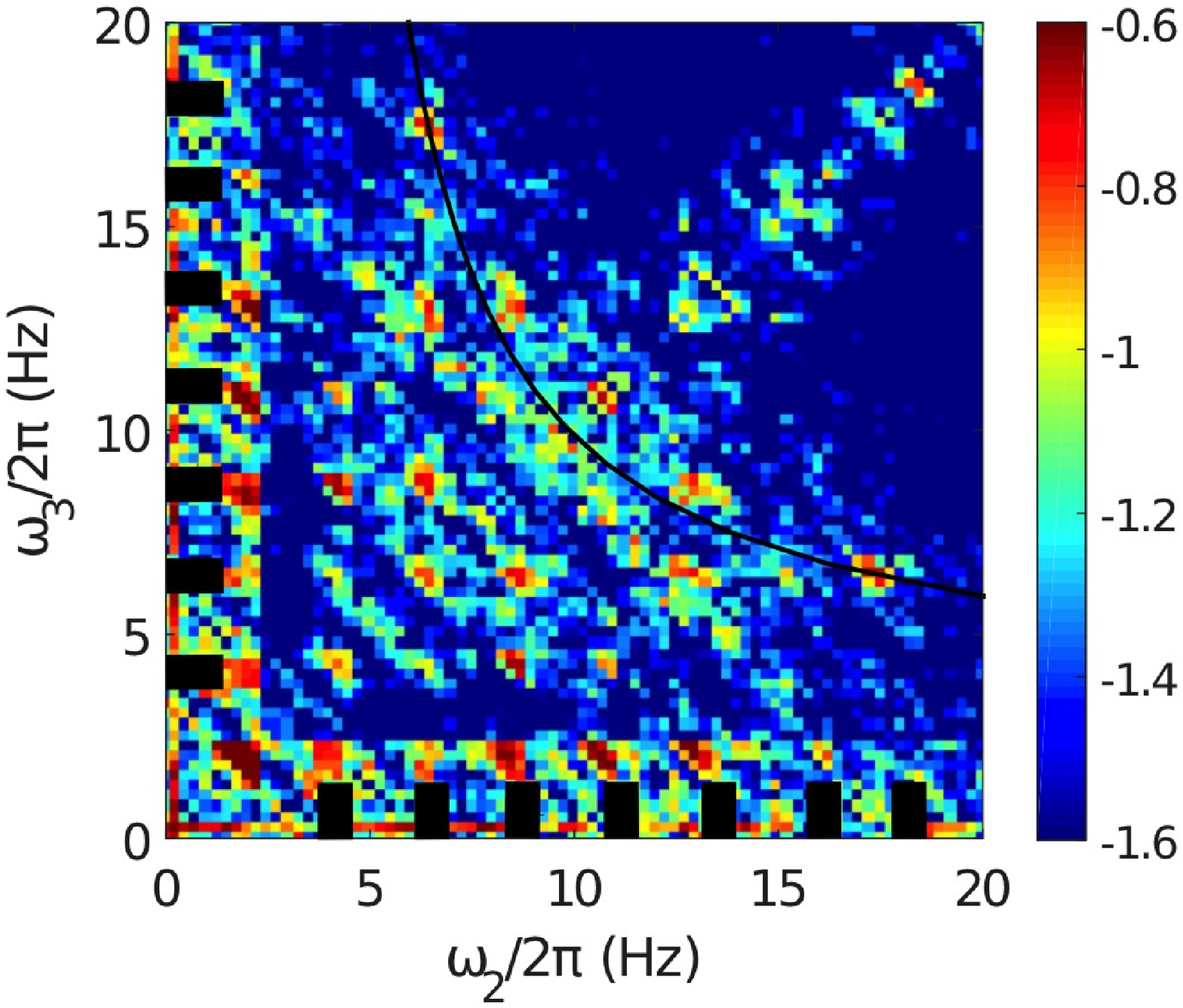}
\caption{Bicoherence of the wave field for experiments carried out in a high aspect ratio channel of size  $1.5$~m $\times$ 10~cm with an rms slope of $3\%$. The black line corresponds to the 1D solutions. (a) full wave field, (b) longitudinal waves (c) transverse waves. The frequencies associated with the discrete transverse modes appear as black rectangle in (c).}
\label{B2brutcanal}
\end{figure}
%Clearer but similar results were obtained in a higher aspect ratio channel of size $1.5$~m $\times$ 10~cm as shown in fig.~\ref{B2brutcanal}. The dotted pattern in the bicoherence of the full wave field (a) is clearly seen to be due to the pure transverse modes (c). The correlations of the longitudinal modes (b) show a continuous pattern. 

We show, in fig.~\ref{B2brutcanal} similar pictures of the bicoherence of the full field as well as the transverse and longitudinal waves in the case of the high aspect ratio channel (the second configuration with a $1.5$~m long channel). The difference between the longitudinal waves and the transverse ones is even more striking in this configuration. The longitunal waves have a continuous spectrum with dominant almost 1D interactions similar to the one reported in the unconfined case at weak nonlinearity. By contrast, the transverse waves show only a few discrete peaks.
%The same type of bicoherence investigations have been carried out in a channel of length 1.50~m and width 10 cm. We plot in fig.~\ref{B2brutcanal}(a) the bicoherence of the full velocity field for experiments with the same central frequency of forcing as used in the vessel and the same rms slope. The figures show even more clearly the strong difference between transverse (c) and longitudinal (b) waves.

All these observations confirm that the energy cascade for discrete turbulence transfers energy essentially from one eigenmode to another but with a whole different scheme from the weak turbulence theory. 

%\begin{figure}[!htb]
%\caption{a)bicoherence of the velocity of the longitudinal waves for experiments carried out in a 1m50 long and 10 cm wide channel with an rms slope of $3\%$. b)bicoherence of the velocity of the transverse waves for experiments carried out in a 1m50 long and 10 cm wide channel with an rms slope of $3\%$. For both a) and b) the red line corresponds to the 1D solutions.}
%\label{B2discretecanal}
%\end{figure}

\section{Discussion}
%%%%%%%%%%%

The weak turbulence theory has been developed for dispersive waves. For a simple dispersion relation $\omega\propto k^\beta$ with $\beta>1$ as is the case for pure capillary waves, only trivial solutions exist for the resonance of unidirectional waves. 1D 3-wave resonances are allowed for non dispersive waves ($\beta=0$) but in that case it could lead to the formation of shocks~\cite{newell_wave_2011}: As all wavetrains travel at the same velocity, they have an infinite time to interact in a ray and exchange energy. For gravity-capillary waves the dispersion relation (\ref{gc_disp}) is more complex and its change of curvature enables non trivial 1D resonances. These waves are dispersive so that the interaction time remains finite but possibly much longer than that of waves that would travel in different directions. Indeed near the gravity-capillary crossover, the waves are only weakly dispersive as a minimum of phase and group velocity is observed. This is most likely the reason why the bicoherence corresponding to 1D resonances is seen to be the strongest in our experiments. Note that the dominance of 1D nonlinear coupling does not mean that 2D interactions are not present. Their coupling is much weaker but most likely they play a major role in the angular redistribution of energy which is clearly operating in our experiments. The total flux of energy is due to subtle global balances of all interactions. Note also that the 2D interactions are more numerous than the 1D interactions so that their role in the global budget can be significant even though individual interactions are much weaker.

Our observations show that due to this peculiar property of gravity-capillary waves, a mixed state of kinetic weak turbulence and discrete weak turbulence can be observed in our channel. Wave turbulence remains continuous in the unconfined direction and discrete in the transverse direction. In the case of strong confinement, the transverse modes become widely separated in frequency so that the linear coupling between modes must be extremely weak. One could then observe several states of weakly coupled kinetic 1D wave turbulence along the unconfined direction but involving distinct transverse modes. The experimental difficulty would be that for a very narrow channel, dissipation by the walls will become very strong and possibly prevent the development of a purely 1D wave turbulence with a single transverse mode.

Our system shares some similarities with nonlinear propagation of light in a multimode fiber although the integrable turbulence observed in optical fibers is very distinct to weak turbulence. Our case is also reminiscent of that of inertial waves of a rotating fluid in a vertically confined channel studied by Scott~\cite{Scott}: Discrete modes are considered in the vertical direction and continuous 2D wave spectrum in the horizontal plane.

\begin{acknowledgements}
%\section{aknowledgements}
This project has received funding from the European Research Council (ERC) under the European Union's Horizon 2020 research and innovation programme (grant agreement No 647018-WATU). We thank Vincent Govart for his technical assistance. We thank Kronos Worldwide, Inc. for kindly providing us with the titanium oxide pigment.
\end{acknowledgements}

%----------------------------------------------------------------------------------------
%	BIBLIOGRAPHY
%\bibliographystyle{unsrt}
\bibliography{mabiblio}

\begin{thebibliography}{43}%
\makeatletter
\providecommand \@ifxundefined [1]{%
 \@ifx{#1\undefined}
}%
\providecommand \@ifnum [1]{%
 \ifnum #1\expandafter \@firstoftwo
 \else \expandafter \@secondoftwo
 \fi
}%
\providecommand \@ifx [1]{%
 \ifx #1\expandafter \@firstoftwo
 \else \expandafter \@secondoftwo
 \fi
}%
\providecommand \natexlab [1]{#1}%
\providecommand \enquote  [1]{``#1''}%
\providecommand \bibnamefont  [1]{#1}%
\providecommand \bibfnamefont [1]{#1}%
\providecommand \citenamefont [1]{#1}%
\providecommand \href@noop [0]{\@secondoftwo}%
\providecommand \href [0]{\begingroup \@sanitize@url \@href}%
\providecommand \@href[1]{\@@startlink{#1}\@@href}%
\providecommand \@@href[1]{\endgroup#1\@@endlink}%
\providecommand \@sanitize@url [0]{\catcode `\\12\catcode `\$12\catcode
  `\&12\catcode `\#12\catcode `\^12\catcode `\_12\catcode `\%12\relax}%
\providecommand \@@startlink[1]{}%
\providecommand \@@endlink[0]{}%
\providecommand \url  [0]{\begingroup\@sanitize@url \@url }%
\providecommand \@url [1]{\endgroup\@href {#1}{\urlprefix }}%
\providecommand \urlprefix  [0]{URL }%
\providecommand \Eprint [0]{\href }%
\providecommand \doibase [0]{http://dx.doi.org/}%
\providecommand \selectlanguage [0]{\@gobble}%
\providecommand \bibinfo  [0]{\@secondoftwo}%
\providecommand \bibfield  [0]{\@secondoftwo}%
\providecommand \translation [1]{[#1]}%
\providecommand \BibitemOpen [0]{}%
\providecommand \bibitemStop [0]{}%
\providecommand \bibitemNoStop [0]{.\EOS\space}%
\providecommand \EOS [0]{\spacefactor3000\relax}%
\providecommand \BibitemShut  [1]{\csname bibitem#1\endcsname}%
\let\auto@bib@innerbib\@empty
%</preamble>
\bibitem [{\citenamefont {Sagdeev}(1979)}]{sagdeev19791976}%
  \BibitemOpen
  \bibfield  {author} {\bibinfo {author} {\bibfnamefont {R.Z.}\ \bibnamefont
  {Sagdeev}},\ }\bibfield  {title} {\enquote {\bibinfo {title} {The 1976
  oppenheimer lectures: Critical problems in plasma astrophysics. i. turbulence
  and nonlinear waves},}\ }\href@noop {} {\bibfield  {journal} {\bibinfo
  {journal} {Rev. Mod. Phys.}\ }\textbf {\bibinfo {volume} {51}},\ \bibinfo
  {pages} {1} (\bibinfo {year} {1979})}\BibitemShut {NoStop}%
\bibitem [{\citenamefont {Picozzi}\ \emph {et~al.}(2014)\citenamefont
  {Picozzi}, \citenamefont {Garnier}, \citenamefont {Hansson}, \citenamefont
  {Suret}, \citenamefont {Randoux}, \citenamefont {Millot},\ and\ \citenamefont
  {Christodoulides}}]{picozzi2014optical}%
  \BibitemOpen
  \bibfield  {author} {\bibinfo {author} {\bibfnamefont {A.}~\bibnamefont
  {Picozzi}}, \bibinfo {author} {\bibfnamefont {J.}~\bibnamefont {Garnier}},
  \bibinfo {author} {\bibfnamefont {T.}~\bibnamefont {Hansson}}, \bibinfo
  {author} {\bibfnamefont {P.}~\bibnamefont {Suret}}, \bibinfo {author}
  {\bibfnamefont {S.}~\bibnamefont {Randoux}}, \bibinfo {author} {\bibfnamefont
  {G.}~\bibnamefont {Millot}}, \ and\ \bibinfo {author} {\bibfnamefont {D.N.}\
  \bibnamefont {Christodoulides}},\ }\bibfield  {title} {\enquote {\bibinfo
  {title} {Optical wave turbulence: Towards a unified nonequilibrium
  thermodynamic formulation of statistical nonlinear optics},}\ }\href@noop {}
  {\bibfield  {journal} {\bibinfo  {journal} {Phys. Rep.}\ }\textbf {\bibinfo
  {volume} {542}},\ \bibinfo {pages} {1--132} (\bibinfo {year}
  {2014})}\BibitemShut {NoStop}%
\bibitem [{\citenamefont {During}\ \emph {et~al.}(2006)\citenamefont {During},
  \citenamefont {Josserand},\ and\ \citenamefont {Rica}}]{During}%
  \BibitemOpen
  \bibfield  {author} {\bibinfo {author} {\bibfnamefont {G.}~\bibnamefont
  {During}}, \bibinfo {author} {\bibfnamefont {C.}~\bibnamefont {Josserand}}, \
  and\ \bibinfo {author} {\bibfnamefont {S.}~\bibnamefont {Rica}},\ }\bibfield
  {title} {\enquote {\bibinfo {title} {{Weak Turbulence for a Vibrating Plate:
  Can One Hear a Kolmogorov Spectrum?}}}\ }\href@noop {} {\bibfield  {journal}
  {\bibinfo  {journal} {Phys. Rev. Lett.}\ }\textbf {\bibinfo {volume} {97}},\
  \bibinfo {pages} {025503} (\bibinfo {year} {2006})}\BibitemShut {NoStop}%
\bibitem [{\citenamefont {Zakharov}\ and\ \citenamefont
  {Filonenko}(1967)}]{Filonenko}%
  \BibitemOpen
  \bibfield  {author} {\bibinfo {author} {\bibfnamefont {V.E.}\ \bibnamefont
  {Zakharov}}\ and\ \bibinfo {author} {\bibfnamefont {N.N.}\ \bibnamefont
  {Filonenko}},\ }\bibfield  {title} {\enquote {\bibinfo {title} {Weak
  turbulence of capillary waves},}\ }\href@noop {} {\bibfield  {journal}
  {\bibinfo  {journal} {J. Appl. Mech. Tech. Phys.}\ }\textbf {\bibinfo
  {volume} {4}},\ \bibinfo {pages} {506} (\bibinfo {year} {1967})}\BibitemShut
  {NoStop}%
\bibitem [{\citenamefont {Hasselmann}(1962)}]{Hasselmann}%
  \BibitemOpen
  \bibfield  {author} {\bibinfo {author} {\bibfnamefont {K.}~\bibnamefont
  {Hasselmann}},\ }\bibfield  {title} {\enquote {\bibinfo {title} {On the
  non-linear energy transfer in gravity-wave spectrum. part 1. general
  theory},}\ }\href@noop {} {\bibfield  {journal} {\bibinfo  {journal} {J.
  Fluid Mech.}\ }\textbf {\bibinfo {volume} {12}},\ \bibinfo {pages} {481--500}
  (\bibinfo {year} {1962})}\BibitemShut {NoStop}%
\bibitem [{\citenamefont {Galtier}(2003)}]{Galtier}%
  \BibitemOpen
  \bibfield  {author} {\bibinfo {author} {\bibfnamefont {S.}~\bibnamefont
  {Galtier}},\ }\bibfield  {title} {\enquote {\bibinfo {title} {{Weak
  inertial-wave turbulence theory}},}\ }\href@noop {} {\bibfield  {journal}
  {\bibinfo  {journal} {Phys. Rev. E}\ }\textbf {\bibinfo {volume} {68}},\
  \bibinfo {pages} {015301} (\bibinfo {year} {2003})}\BibitemShut {NoStop}%
\bibitem [{\citenamefont {Nazarenko}(2011)}]{Nazarenko}%
  \BibitemOpen
  \bibfield  {author} {\bibinfo {author} {\bibfnamefont {S.}~\bibnamefont
  {Nazarenko}},\ }\href@noop {} {\emph {\bibinfo {title} {{Wave
  turbulence}}}},\ Vol.\ \bibinfo {volume} {825}\ (\bibinfo  {publisher}
  {Springer Science {\&} Business Media},\ \bibinfo {year} {2011})\BibitemShut
  {NoStop}%
\bibitem [{\citenamefont {Newell}\ and\ \citenamefont
  {Rumpf}(2011)}]{newell_wave_2011}%
  \BibitemOpen
  \bibfield  {author} {\bibinfo {author} {\bibfnamefont {A.C.}\ \bibnamefont
  {Newell}}\ and\ \bibinfo {author} {\bibfnamefont {B.}~\bibnamefont {Rumpf}},\
  }\bibfield  {title} {\enquote {\bibinfo {title} {Wave {Turbulence}},}\ }\href
  {\doibase 10.1146/annurev-fluid-122109-160807} {\bibfield  {journal}
  {\bibinfo  {journal} {Ann. Rev. Fluid Mech.}\ }\textbf {\bibinfo {volume}
  {43}},\ \bibinfo {pages} {59--78} (\bibinfo {year} {2011})}\BibitemShut
  {NoStop}%
\bibitem [{\citenamefont {Zakharov}\ \emph {et~al.}(1992)\citenamefont
  {Zakharov}, \citenamefont {L'vov},\ and\ \citenamefont {Falkovich}}]{R1}%
  \BibitemOpen
  \bibfield  {author} {\bibinfo {author} {\bibfnamefont {V.~E.}\ \bibnamefont
  {Zakharov}}, \bibinfo {author} {\bibfnamefont {V.~S.}\ \bibnamefont {L'vov}},
  \ and\ \bibinfo {author} {\bibfnamefont {G.}~\bibnamefont {Falkovich}},\
  }\href@noop {} {\emph {\bibinfo {title} {Kolmogorov Spectra of Turbulence}}}\
  (\bibinfo  {publisher} {Springer},\ \bibinfo {address} {Berlin},\ \bibinfo
  {year} {1992})\BibitemShut {NoStop}%
\bibitem [{\citenamefont {Aubourg}\ and\ \citenamefont
  {Mordant}(2015)}]{aubourg_nonlocal_2015}%
  \BibitemOpen
  \bibfield  {author} {\bibinfo {author} {\bibfnamefont {Q.}~\bibnamefont
  {Aubourg}}\ and\ \bibinfo {author} {\bibfnamefont {N.}~\bibnamefont
  {Mordant}},\ }\bibfield  {title} {\enquote {\bibinfo {title} {Nonlocal
  resonances in weak turbulence of gravity-capillary waves},}\ }\href@noop {}
  {\bibfield  {journal} {\bibinfo  {journal} {Phys. Rev. Lett.}\ }\textbf
  {\bibinfo {volume} {114}},\ \bibinfo {pages} {144501} (\bibinfo {year}
  {2015})}\BibitemShut {NoStop}%
\bibitem [{\citenamefont {Aubourg}\ and\ \citenamefont
  {Mordant}(2016)}]{aubourg2016investigation}%
  \BibitemOpen
  \bibfield  {author} {\bibinfo {author} {\bibfnamefont {Q.}~\bibnamefont
  {Aubourg}}\ and\ \bibinfo {author} {\bibfnamefont {N.}~\bibnamefont
  {Mordant}},\ }\bibfield  {title} {\enquote {\bibinfo {title} {Investigation
  of resonances in gravity-capillary wave turbulence},}\ }\href@noop {}
  {\bibfield  {journal} {\bibinfo  {journal} {Phys. Rev. Fluids}\ }\textbf
  {\bibinfo {volume} {1}},\ \bibinfo {pages} {023701} (\bibinfo {year}
  {2016})}\BibitemShut {NoStop}%
\bibitem [{\citenamefont {Kartashova}(1994)}]{Kartashova:1994}%
  \BibitemOpen
  \bibfield  {author} {\bibinfo {author} {\bibfnamefont {E.}~\bibnamefont
  {Kartashova}},\ }\bibfield  {title} {\enquote {\bibinfo {title} {Weakly
  nonlinear theory of finite size effects in resonators},}\ }\href@noop {}
  {\bibfield  {journal} {\bibinfo  {journal} {Phys. Rev. Lett.}\ }\textbf
  {\bibinfo {volume} {72}},\ \bibinfo {pages} {2013--2016} (\bibinfo {year}
  {1994})}\BibitemShut {NoStop}%
\bibitem [{\citenamefont {Kartashova}(2009)}]{kartashova2009discrete}%
  \BibitemOpen
  \bibfield  {author} {\bibinfo {author} {\bibfnamefont {E.}~\bibnamefont
  {Kartashova}},\ }\bibfield  {title} {\enquote {\bibinfo {title} {Discrete
  wave turbulence},}\ }\href@noop {} {\bibfield  {journal} {\bibinfo  {journal}
  {EPL}\ }\textbf {\bibinfo {volume} {87}},\ \bibinfo {pages} {44001} (\bibinfo
  {year} {2009})}\BibitemShut {NoStop}%
\bibitem [{\citenamefont {Pan}\ and\ \citenamefont {Yue}(2017)}]{Pan}%
  \BibitemOpen
  \bibfield  {author} {\bibinfo {author} {\bibfnamefont {Yulin}\ \bibnamefont
  {Pan}}\ and\ \bibinfo {author} {\bibfnamefont {Dick K~P}\ \bibnamefont
  {Yue}},\ }\bibfield  {title} {\enquote {\bibinfo {title} {{Understanding
  discrete capillary-wave turbulence using a quasi-resonant kinetic
  equation}},}\ }\href@noop {} {\bibfield  {journal} {\bibinfo  {journal}
  {Journal Of Fluid Mechanics}\ }\textbf {\bibinfo {volume} {816}},\ \bibinfo
  {pages} {96} (\bibinfo {year} {2017})}\BibitemShut {NoStop}%
\bibitem [{\citenamefont {L'vov}\ and\ \citenamefont
  {Nazarenko}(2010)}]{l2010discrete}%
  \BibitemOpen
  \bibfield  {author} {\bibinfo {author} {\bibfnamefont {V.S.}\ \bibnamefont
  {L'vov}}\ and\ \bibinfo {author} {\bibfnamefont {S.}~\bibnamefont
  {Nazarenko}},\ }\bibfield  {title} {\enquote {\bibinfo {title} {Discrete and
  mesoscopic regimes of finite-size wave turbulence},}\ }\href@noop {}
  {\bibfield  {journal} {\bibinfo  {journal} {Phys. Rev. E}\ }\textbf {\bibinfo
  {volume} {82}},\ \bibinfo {pages} {056322} (\bibinfo {year}
  {2010})}\BibitemShut {NoStop}%
\bibitem [{\citenamefont {Denissenko}\ \emph {et~al.}(2007)\citenamefont
  {Denissenko}, \citenamefont {Lukaschuk},\ and\ \citenamefont
  {Nazarenko}}]{denissenko2007gravity}%
  \BibitemOpen
  \bibfield  {author} {\bibinfo {author} {\bibfnamefont {P.}~\bibnamefont
  {Denissenko}}, \bibinfo {author} {\bibfnamefont {S.}~\bibnamefont
  {Lukaschuk}}, \ and\ \bibinfo {author} {\bibfnamefont {S.}~\bibnamefont
  {Nazarenko}},\ }\bibfield  {title} {\enquote {\bibinfo {title} {Gravity wave
  turbulence in a laboratory flume},}\ }\href@noop {} {\bibfield  {journal}
  {\bibinfo  {journal} {Phys. Rev. Lett.}\ }\textbf {\bibinfo {volume} {99}},\
  \bibinfo {pages} {014501} (\bibinfo {year} {2007})}\BibitemShut {NoStop}%
\bibitem [{\citenamefont {Nazarenko}\ \emph {et~al.}(2010)\citenamefont
  {Nazarenko}, \citenamefont {Lukaschuk}, \citenamefont {McLelland},\ and\
  \citenamefont {Denissenko}}]{nazarenko2010statistics}%
  \BibitemOpen
  \bibfield  {author} {\bibinfo {author} {\bibfnamefont {S.}~\bibnamefont
  {Nazarenko}}, \bibinfo {author} {\bibfnamefont {S.}~\bibnamefont
  {Lukaschuk}}, \bibinfo {author} {\bibfnamefont {S.}~\bibnamefont
  {McLelland}}, \ and\ \bibinfo {author} {\bibfnamefont {P.}~\bibnamefont
  {Denissenko}},\ }\bibfield  {title} {\enquote {\bibinfo {title} {Statistics
  of surface gravity wave turbulence in the space and time domains},}\
  }\href@noop {} {\bibfield  {journal} {\bibinfo  {journal} {J. Fluid Mech.}\
  }\textbf {\bibinfo {volume} {642}},\ \bibinfo {pages} {395--420} (\bibinfo
  {year} {2010})}\BibitemShut {NoStop}%
\bibitem [{\citenamefont {Nazarenko}(2006)}]{nazarenko2006sandpile}%
  \BibitemOpen
  \bibfield  {author} {\bibinfo {author} {\bibfnamefont {S.}~\bibnamefont
  {Nazarenko}},\ }\bibfield  {title} {\enquote {\bibinfo {title} {Sandpile
  behaviour in discrete water-wave turbulence},}\ }\href@noop {} {\bibfield
  {journal} {\bibinfo  {journal} {J. Stat. Mech.}\ }\textbf {\bibinfo {volume}
  {2006}},\ \bibinfo {pages} {L02002} (\bibinfo {year} {2006})}\BibitemShut
  {NoStop}%
\bibitem [{\citenamefont {Pushkarev}(1999)}]{pushkarev1999kolmogorov}%
  \BibitemOpen
  \bibfield  {author} {\bibinfo {author} {\bibfnamefont {A.N.}\ \bibnamefont
  {Pushkarev}},\ }\bibfield  {title} {\enquote {\bibinfo {title} {On the
  kolmogorov and frozen turbulence in numerical simulation of capillary
  waves},}\ }\href@noop {} {\bibfield  {journal} {\bibinfo  {journal} {Eur. J.
  Mech. B}\ }\textbf {\bibinfo {volume} {18}},\ \bibinfo {pages} {345--351}
  (\bibinfo {year} {1999})}\BibitemShut {NoStop}%
\bibitem [{\citenamefont {Falcon}\ \emph {et~al.}(2007)\citenamefont {Falcon},
  \citenamefont {Laroche},\ and\ \citenamefont
  {Fauve}}]{falcon2007observation}%
  \BibitemOpen
  \bibfield  {author} {\bibinfo {author} {\bibfnamefont {E.}~\bibnamefont
  {Falcon}}, \bibinfo {author} {\bibfnamefont {C.}~\bibnamefont {Laroche}}, \
  and\ \bibinfo {author} {\bibfnamefont {S.}~\bibnamefont {Fauve}},\ }\bibfield
   {title} {\enquote {\bibinfo {title} {Observation of gravity-capillary wave
  turbulence},}\ }\href@noop {} {\bibfield  {journal} {\bibinfo  {journal}
  {Phys. Rev. Lett.}\ }\textbf {\bibinfo {volume} {98}},\ \bibinfo {pages}
  {094503} (\bibinfo {year} {2007})}\BibitemShut {NoStop}%
\bibitem [{\citenamefont {Aubourg}\ \emph {et~al.}(2017)\citenamefont
  {Aubourg}, \citenamefont {A.}, \citenamefont {C.}, \citenamefont {F.},
  \citenamefont {J.}, \citenamefont {S.},\ and\ \citenamefont
  {N.}}]{Aubourg2017}%
  \BibitemOpen
  \bibfield  {author} {\bibinfo {author} {\bibfnamefont {Q.}~\bibnamefont
  {Aubourg}}, \bibinfo {author} {\bibfnamefont {Campagne}\ \bibnamefont {A.}},
  \bibinfo {author} {\bibfnamefont {Peureux}\ \bibnamefont {C.}}, \bibinfo
  {author} {\bibfnamefont {Ardhuin}\ \bibnamefont {F.}}, \bibinfo {author}
  {\bibfnamefont {Sommeria}\ \bibnamefont {J.}}, \bibinfo {author}
  {\bibfnamefont {Viboud}\ \bibnamefont {S.}}, \ and\ \bibinfo {author}
  {\bibfnamefont {Mordant}\ \bibnamefont {N.}},\ }\bibfield  {title} {\enquote
  {\bibinfo {title} {Three-wave and four-wave interactions in gravity wave
  turbulence},}\ }\href@noop {} {\bibfield  {journal} {\bibinfo  {journal}
  {Phys. Rev. Fluids}\ }\textbf {\bibinfo {volume} {2}},\ \bibinfo {pages}
  {114802} (\bibinfo {year} {2017})}\BibitemShut {NoStop}%
\bibitem [{\citenamefont {Leckler}\ \emph {et~al.}(2015)\citenamefont
  {Leckler}, \citenamefont {Ardhuin}, \citenamefont {Peureux}, \citenamefont
  {Benetazzo}, \citenamefont {Bergamasco},\ and\ \citenamefont
  {Dulov}}]{Leckler}%
  \BibitemOpen
  \bibfield  {author} {\bibinfo {author} {\bibfnamefont {F.}~\bibnamefont
  {Leckler}}, \bibinfo {author} {\bibfnamefont {F.}~\bibnamefont {Ardhuin}},
  \bibinfo {author} {\bibfnamefont {C.}~\bibnamefont {Peureux}}, \bibinfo
  {author} {\bibfnamefont {A.}~\bibnamefont {Benetazzo}}, \bibinfo {author}
  {\bibfnamefont {F.}~\bibnamefont {Bergamasco}}, \ and\ \bibinfo {author}
  {\bibfnamefont {V.}~\bibnamefont {Dulov}},\ }\bibfield  {title} {\enquote
  {\bibinfo {title} {{Analysis and Interpretation of Frequency-Wavenumber
  Spectra of Young Wind Waves}},}\ }\href {\doibase 10.1175/JPO-D-14-0237.1}
  {\bibfield  {journal} {\bibinfo  {journal} {J. Phys. Ocean.}\ }\textbf
  {\bibinfo {volume} {45}},\ \bibinfo {pages} {2484----2496} (\bibinfo {year}
  {2015})}\BibitemShut {NoStop}%
\bibitem [{\citenamefont {Lenain}\ and\ \citenamefont
  {Melville}(2017)}]{Lenain}%
  \BibitemOpen
  \bibfield  {author} {\bibinfo {author} {\bibfnamefont {L.}~\bibnamefont
  {Lenain}}\ and\ \bibinfo {author} {\bibfnamefont {W.~K.}\ \bibnamefont
  {Melville}},\ }\bibfield  {title} {\enquote {\bibinfo {title} {Measurements
  of the directional spectrum across the equilibrium saturation ranges of
  wind-generated surface waves.}}\ }\href@noop {} {\bibfield  {journal}
  {\bibinfo  {journal} {J. Phys. Ocean.}\ }\textbf {\bibinfo {volume} {47}},\
  \bibinfo {pages} {2123-- 2138} (\bibinfo {year} {2017})}\BibitemShut
  {NoStop}%
\bibitem [{\citenamefont {Wright}\ \emph {et~al.}(1997)\citenamefont {Wright},
  \citenamefont {Budakian}, \citenamefont {Pine},\ and\ \citenamefont
  {Putterman}}]{wright1997imaging}%
  \BibitemOpen
  \bibfield  {author} {\bibinfo {author} {\bibfnamefont {W.B.}\ \bibnamefont
  {Wright}}, \bibinfo {author} {\bibfnamefont {R.}~\bibnamefont {Budakian}},
  \bibinfo {author} {\bibfnamefont {D.J.}\ \bibnamefont {Pine}}, \ and\
  \bibinfo {author} {\bibfnamefont {S.J.}\ \bibnamefont {Putterman}},\
  }\bibfield  {title} {\enquote {\bibinfo {title} {Imaging of intermittency in
  ripple-wave turbulence},}\ }\href@noop {} {\bibfield  {journal} {\bibinfo
  {journal} {Science}\ }\textbf {\bibinfo {volume} {278}},\ \bibinfo {pages}
  {1609--1612} (\bibinfo {year} {1997})}\BibitemShut {NoStop}%
\bibitem [{\citenamefont {Deike}\ \emph {et~al.}(2012)\citenamefont {Deike},
  \citenamefont {Berhanu},\ and\ \citenamefont {Falcon}}]{deike2012decay}%
  \BibitemOpen
  \bibfield  {author} {\bibinfo {author} {\bibfnamefont {L.}~\bibnamefont
  {Deike}}, \bibinfo {author} {\bibfnamefont {M.}~\bibnamefont {Berhanu}}, \
  and\ \bibinfo {author} {\bibfnamefont {E.}~\bibnamefont {Falcon}},\
  }\bibfield  {title} {\enquote {\bibinfo {title} {Decay of capillary wave
  turbulence},}\ }\href@noop {} {\bibfield  {journal} {\bibinfo  {journal}
  {Phys. Rev. E}\ }\textbf {\bibinfo {volume} {85}},\ \bibinfo {pages} {066311}
  (\bibinfo {year} {2012})}\BibitemShut {NoStop}%
\bibitem [{\citenamefont {Kolmakov}\ \emph {et~al.}(2004)\citenamefont
  {Kolmakov}, \citenamefont {Levchenko}, \citenamefont {Brazhnikov},
  \citenamefont {Mezhov-Deglin}, \citenamefont {Silchenko},\ and\ \citenamefont
  {Mcclintock}}]{Kolmakov}%
  \BibitemOpen
  \bibfield  {author} {\bibinfo {author} {\bibfnamefont {G~V}\ \bibnamefont
  {Kolmakov}}, \bibinfo {author} {\bibfnamefont {A~A}\ \bibnamefont
  {Levchenko}}, \bibinfo {author} {\bibfnamefont {M}~\bibnamefont
  {Brazhnikov}}, \bibinfo {author} {\bibfnamefont {L~P}\ \bibnamefont
  {Mezhov-Deglin}}, \bibinfo {author} {\bibfnamefont {A}~\bibnamefont
  {Silchenko}}, \ and\ \bibinfo {author} {\bibfnamefont {P}~\bibnamefont
  {Mcclintock}},\ }\bibfield  {title} {\enquote {\bibinfo {title}
  {{Quasiadiabatic Decay of Capillary Turbulence on the Charged Surface of
  Liquid Hydrogen}},}\ }\href@noop {} {\bibfield  {journal} {\bibinfo
  {journal} {Physical Review Letters}\ }\textbf {\bibinfo {volume} {93}},\
  \bibinfo {pages} {074501} (\bibinfo {year} {2004})}\BibitemShut {NoStop}%
\bibitem [{\citenamefont {Xia}\ \emph {et~al.}(2010)\citenamefont {Xia},
  \citenamefont {Shats},\ and\ \citenamefont {Punzmann}}]{Xia}%
  \BibitemOpen
  \bibfield  {author} {\bibinfo {author} {\bibfnamefont {H}~\bibnamefont
  {Xia}}, \bibinfo {author} {\bibfnamefont {M~G}\ \bibnamefont {Shats}}, \ and\
  \bibinfo {author} {\bibfnamefont {H}~\bibnamefont {Punzmann}},\ }\bibfield
  {title} {\enquote {\bibinfo {title} {{Modulation instability and capillary
  wave turbulence}},}\ }\href@noop {} {\bibfield  {journal} {\bibinfo
  {journal} {Epl}\ }\textbf {\bibinfo {volume} {91}},\ \bibinfo {pages} {14002}
  (\bibinfo {year} {2010})}\BibitemShut {NoStop}%
\bibitem [{\citenamefont {Punzmann}\ \emph {et~al.}(2009)\citenamefont
  {Punzmann}, \citenamefont {Shats},\ and\ \citenamefont {Xia}}]{Punzmann}%
  \BibitemOpen
  \bibfield  {author} {\bibinfo {author} {\bibfnamefont {H}~\bibnamefont
  {Punzmann}}, \bibinfo {author} {\bibfnamefont {M~G}\ \bibnamefont {Shats}}, \
  and\ \bibinfo {author} {\bibfnamefont {H}~\bibnamefont {Xia}},\ }\bibfield
  {title} {\enquote {\bibinfo {title} {{Phase Randomization of Three-Wave
  Interactions in Capillary Waves}},}\ }\href@noop {} {\bibfield  {journal}
  {\bibinfo  {journal} {Physical Review Letters}\ }\textbf {\bibinfo {volume}
  {103}},\ \bibinfo {pages} {064502} (\bibinfo {year} {2009})}\BibitemShut
  {NoStop}%
\bibitem [{\citenamefont {Campagne}\ \emph {et~al.}(2018)\citenamefont
  {Campagne}, \citenamefont {Hassaini}, \citenamefont {Redor}, \citenamefont
  {Sommeria}, \citenamefont {Valran}, \citenamefont {Viboud},\ and\
  \citenamefont {Mordant}}]{Campagne}%
  \BibitemOpen
  \bibfield  {author} {\bibinfo {author} {\bibfnamefont {A.}~\bibnamefont
  {Campagne}}, \bibinfo {author} {\bibfnamefont {R.}~\bibnamefont {Hassaini}},
  \bibinfo {author} {\bibfnamefont {I.}~\bibnamefont {Redor}}, \bibinfo
  {author} {\bibfnamefont {J.}~\bibnamefont {Sommeria}}, \bibinfo {author}
  {\bibfnamefont {T.}~\bibnamefont {Valran}}, \bibinfo {author} {\bibfnamefont
  {S.}~\bibnamefont {Viboud}}, \ and\ \bibinfo {author} {\bibfnamefont
  {N.}~\bibnamefont {Mordant}},\ }\bibfield  {title} {\enquote {\bibinfo
  {title} {Impact of dissipation on the energy spectrum of experimental
  turbulence of gravity surface waves},}\ }\href@noop {} {\bibfield  {journal}
  {\bibinfo  {journal} {Phys. Rev. Fluids}\ }\textbf {\bibinfo {volume} {3}},\
  \bibinfo {pages} {044801} (\bibinfo {year} {2018})}\BibitemShut {NoStop}%
\bibitem [{\citenamefont {Miquel}\ \emph {et~al.}(2014)\citenamefont {Miquel},
  \citenamefont {Alexakis},\ and\ \citenamefont {Mordant}}]{R23}%
  \BibitemOpen
  \bibfield  {author} {\bibinfo {author} {\bibfnamefont {B.}~\bibnamefont
  {Miquel}}, \bibinfo {author} {\bibfnamefont {A.}~\bibnamefont {Alexakis}}, \
  and\ \bibinfo {author} {\bibfnamefont {N.}~\bibnamefont {Mordant}},\
  }\bibfield  {title} {\enquote {\bibinfo {title} {Role of dissipation in
  flexural wave turbulence: from experimental spectrum to kolmogorov-zakharov
  spectrum},}\ }\href@noop {} {\bibfield  {journal} {\bibinfo  {journal} {Phys.
  Rev. E}\ }\textbf {\bibinfo {volume} {89}},\ \bibinfo {pages} {062925}
  (\bibinfo {year} {2014})}\BibitemShut {NoStop}%
\bibitem [{\citenamefont {Humbert}\ \emph {et~al.}(2013)\citenamefont
  {Humbert}, \citenamefont {Cadot}, \citenamefont {D\"uring}, \citenamefont
  {Josserand}, \citenamefont {Rica},\ and\ \citenamefont {Touz\'e}}]{Humbert}%
  \BibitemOpen
  \bibfield  {author} {\bibinfo {author} {\bibfnamefont {T.}~\bibnamefont
  {Humbert}}, \bibinfo {author} {\bibfnamefont {O.}~\bibnamefont {Cadot}},
  \bibinfo {author} {\bibfnamefont {G.}~\bibnamefont {D\"uring}}, \bibinfo
  {author} {\bibfnamefont {C.}~\bibnamefont {Josserand}}, \bibinfo {author}
  {\bibfnamefont {S.}~\bibnamefont {Rica}}, \ and\ \bibinfo {author}
  {\bibfnamefont {C.}~\bibnamefont {Touz\'e}},\ }\bibfield  {title} {\enquote
  {\bibinfo {title} {Wave turbulence in vibrating plates : the effect of
  damping},}\ }\href@noop {} {\bibfield  {journal} {\bibinfo  {journal} {EPL}\
  }\textbf {\bibinfo {volume} {102}},\ \bibinfo {pages} {30002} (\bibinfo
  {year} {2013})}\BibitemShut {NoStop}%
\bibitem [{\citenamefont {Deike}\ \emph {et~al.}(2014)\citenamefont {Deike},
  \citenamefont {berhanu},\ and\ \citenamefont {Falcon}}]{Deike:2014kt}%
  \BibitemOpen
  \bibfield  {author} {\bibinfo {author} {\bibfnamefont {L}~\bibnamefont
  {Deike}}, \bibinfo {author} {\bibfnamefont {Michael}\ \bibnamefont
  {berhanu}}, \ and\ \bibinfo {author} {\bibfnamefont {Eric}\ \bibnamefont
  {Falcon}},\ }\bibfield  {title} {\enquote {\bibinfo {title} {{Energy flux
  measurement from the dissipated energy in capillary wave turbulence}},}\
  }\href@noop {} {\bibfield  {journal} {\bibinfo  {journal} {Physical Review
  E}\ }\textbf {\bibinfo {volume} {89}},\ \bibinfo {pages} {881} (\bibinfo
  {year} {2014})}\BibitemShut {NoStop}%
\bibitem [{\citenamefont {Przadka}\ \emph {et~al.}(2011)\citenamefont
  {Przadka}, \citenamefont {Cabane}, \citenamefont {Pagneux}, \citenamefont
  {Maurel},\ and\ \citenamefont {Petitjeans}}]{Przadka}%
  \BibitemOpen
  \bibfield  {author} {\bibinfo {author} {\bibfnamefont {A.}~\bibnamefont
  {Przadka}}, \bibinfo {author} {\bibfnamefont {B.}~\bibnamefont {Cabane}},
  \bibinfo {author} {\bibfnamefont {V.}~\bibnamefont {Pagneux}}, \bibinfo
  {author} {\bibfnamefont {A.}~\bibnamefont {Maurel}}, \ and\ \bibinfo {author}
  {\bibfnamefont {P.}~\bibnamefont {Petitjeans}},\ }\bibfield  {title}
  {\enquote {\bibinfo {title} {{Fourier transform profilometry for water waves:
  how to achieve clean water attenuation with diffusive reflection at the water
  surface?}}}\ }\href@noop {} {\bibfield  {journal} {\bibinfo  {journal}
  {Experiments In Fluids}\ }\textbf {\bibinfo {volume} {52}},\ \bibinfo {pages}
  {519--527} (\bibinfo {year} {2011})}\BibitemShut {NoStop}%
\bibitem [{\citenamefont {Cobelli}\ \emph {et~al.}(2009)\citenamefont
  {Cobelli}, \citenamefont {Maurel}, \citenamefont {Pagneux},\ and\
  \citenamefont {Petitjeans}}]{cobelli_global_2009}%
  \BibitemOpen
  \bibfield  {author} {\bibinfo {author} {\bibfnamefont {P.~J.}\ \bibnamefont
  {Cobelli}}, \bibinfo {author} {\bibfnamefont {A.}~\bibnamefont {Maurel}},
  \bibinfo {author} {\bibfnamefont {V.}~\bibnamefont {Pagneux}}, \ and\
  \bibinfo {author} {\bibfnamefont {P.}~\bibnamefont {Petitjeans}},\ }\bibfield
   {title} {\enquote {\bibinfo {title} {Global measurement of water waves by
  {Fourier} transform profilometry},}\ }\href {\doibase
  10.1007/s00348-009-0611-z} {\bibfield  {journal} {\bibinfo  {journal} {Exp.
  Fluids}\ }\textbf {\bibinfo {volume} {46}},\ \bibinfo {pages} {1037--1047}
  (\bibinfo {year} {2009})}\BibitemShut {NoStop}%
\bibitem [{\citenamefont {Maurel}\ \emph {et~al.}(2009)\citenamefont {Maurel},
  \citenamefont {Cobelli}, \citenamefont {Pagneux},\ and\ \citenamefont
  {Petitjeans}}]{maurel_experimental_2009}%
  \BibitemOpen
  \bibfield  {author} {\bibinfo {author} {\bibfnamefont {A.}~\bibnamefont
  {Maurel}}, \bibinfo {author} {\bibfnamefont {P.}~\bibnamefont {Cobelli}},
  \bibinfo {author} {\bibfnamefont {V.}~\bibnamefont {Pagneux}}, \ and\
  \bibinfo {author} {\bibfnamefont {P.}~\bibnamefont {Petitjeans}},\ }\bibfield
   {title} {\enquote {\bibinfo {title} {Experimental and theoretical inspection
  of the phase-to-height relation in {Fourier} transform profilometry},}\
  }\href@noop {} {\bibfield  {journal} {\bibinfo  {journal} {Applied Optics}\
  }\textbf {\bibinfo {volume} {48}},\ \bibinfo {pages} {380--392} (\bibinfo
  {year} {2009})}\BibitemShut {NoStop}%
\bibitem [{\citenamefont {Cobelli}\ \emph {et~al.}(2011)\citenamefont
  {Cobelli}, \citenamefont {Przadka}, \citenamefont {Petitjeans}, \citenamefont
  {Lagubeau}, \citenamefont {Pagneux},\ and\ \citenamefont {Maurel}}]{Cobelli}%
  \BibitemOpen
  \bibfield  {author} {\bibinfo {author} {\bibfnamefont {P.}~\bibnamefont
  {Cobelli}}, \bibinfo {author} {\bibfnamefont {A.}~\bibnamefont {Przadka}},
  \bibinfo {author} {\bibfnamefont {P.}~\bibnamefont {Petitjeans}}, \bibinfo
  {author} {\bibfnamefont {G.}~\bibnamefont {Lagubeau}}, \bibinfo {author}
  {\bibfnamefont {V.}~\bibnamefont {Pagneux}}, \ and\ \bibinfo {author}
  {\bibfnamefont {A.}~\bibnamefont {Maurel}},\ }\bibfield  {title} {\enquote
  {\bibinfo {title} {{Different Regimes for Water Wave Turbulence}},}\
  }\href@noop {} {\bibfield  {journal} {\bibinfo  {journal} {Phys. Rev. Lett.}\
  }\textbf {\bibinfo {volume} {107}},\ \bibinfo {pages} {214503} (\bibinfo
  {year} {2011})}\BibitemShut {NoStop}%
\bibitem [{\citenamefont {Berhanu}\ \emph {et~al.}(2018)\citenamefont
  {Berhanu}, \citenamefont {Falcon},\ and\ \citenamefont {Deike}}]{Berhanu2}%
  \BibitemOpen
  \bibfield  {author} {\bibinfo {author} {\bibfnamefont {M.}~\bibnamefont
  {Berhanu}}, \bibinfo {author} {\bibfnamefont {E.}~\bibnamefont {Falcon}}, \
  and\ \bibinfo {author} {\bibfnamefont {L.}~\bibnamefont {Deike}},\ }\bibfield
   {title} {\enquote {\bibinfo {title} {Turbulence of capillary waves forced by
  steep gravity waves},}\ }\href@noop {} {\bibfield  {journal} {\bibinfo
  {journal} {submitted to J. Fluid Mech.}\ } (\bibinfo {year}
  {2018})}\BibitemShut {NoStop}%
\bibitem [{\citenamefont {Hassaini}\ and\ \citenamefont
  {Mordant}(2017)}]{hassaini2017transition}%
  \BibitemOpen
  \bibfield  {author} {\bibinfo {author} {\bibfnamefont {R.}~\bibnamefont
  {Hassaini}}\ and\ \bibinfo {author} {\bibfnamefont {N.}~\bibnamefont
  {Mordant}},\ }\bibfield  {title} {\enquote {\bibinfo {title} {Transition from
  weak wave turbulence to soliton gas},}\ }\href@noop {} {\bibfield  {journal}
  {\bibinfo  {journal} {Phys. Rev. Fluids}\ }\textbf {\bibinfo {volume} {2}},\
  \bibinfo {pages} {094803} (\bibinfo {year} {2017})}\BibitemShut {NoStop}%
\bibitem [{\citenamefont {Berhanu}\ and\ \citenamefont
  {Falcon}(2013)}]{Berhanu}%
  \BibitemOpen
  \bibfield  {author} {\bibinfo {author} {\bibfnamefont {M.}~\bibnamefont
  {Berhanu}}\ and\ \bibinfo {author} {\bibfnamefont {E.}~\bibnamefont
  {Falcon}},\ }\bibfield  {title} {\enquote {\bibinfo {title}
  {{Space-time-resolved capillary wave turbulence}},}\ }\href@noop {}
  {\bibfield  {journal} {\bibinfo  {journal} {Phys. Rev. E}\ }\textbf {\bibinfo
  {volume} {89}},\ \bibinfo {pages} {033003} (\bibinfo {year}
  {2013})}\BibitemShut {NoStop}%
\bibitem [{\citenamefont {Deike}\ \emph {et~al.}(2015)\citenamefont {Deike},
  \citenamefont {Miquel}, \citenamefont {Gutierrez}, \citenamefont {Jamin},
  \citenamefont {Semin}, \citenamefont {Berhanu}, \citenamefont {Falcon},\ and\
  \citenamefont {Bonnefoy}}]{Deike}%
  \BibitemOpen
  \bibfield  {author} {\bibinfo {author} {\bibfnamefont {L.}~\bibnamefont
  {Deike}}, \bibinfo {author} {\bibfnamefont {B.}~\bibnamefont {Miquel}},
  \bibinfo {author} {\bibfnamefont {P.}~\bibnamefont {Gutierrez}}, \bibinfo
  {author} {\bibfnamefont {T.}~\bibnamefont {Jamin}}, \bibinfo {author}
  {\bibfnamefont {B.}~\bibnamefont {Semin}}, \bibinfo {author} {\bibfnamefont
  {M.}~\bibnamefont {Berhanu}}, \bibinfo {author} {\bibfnamefont
  {E.}~\bibnamefont {Falcon}}, \ and\ \bibinfo {author} {\bibfnamefont
  {F.}~\bibnamefont {Bonnefoy}},\ }\bibfield  {title} {\enquote {\bibinfo
  {title} {Role of the basin boundary conditions in gravity wave turbulence},}\
  }\href {\doibase 10.1017/jfm.2015.494} {\bibfield  {journal} {\bibinfo
  {journal} {J. Fluid Mech.}\ }\textbf {\bibinfo {volume} {781}},\ \bibinfo
  {pages} {196--225} (\bibinfo {year} {2015})}\BibitemShut {NoStop}%
\bibitem [{\citenamefont {Mordant}(2010)}]{R21}%
  \BibitemOpen
  \bibfield  {author} {\bibinfo {author} {\bibfnamefont {N.}~\bibnamefont
  {Mordant}},\ }\bibfield  {title} {\enquote {\bibinfo {title} {Fourier
  analysis of wave turbulence in a thin elastic plate},}\ }\href@noop {}
  {\bibfield  {journal} {\bibinfo  {journal} {Eur. Phys. J. B}\ }\textbf
  {\bibinfo {volume} {76}},\ \bibinfo {pages} {537--545} (\bibinfo {year}
  {2010})}\BibitemShut {NoStop}%
\bibitem [{\citenamefont {Aubourg}(2016)}]{theseAubourg2016}%
  \BibitemOpen
  \bibfield  {author} {\bibinfo {author} {\bibfnamefont {Q.}~\bibnamefont
  {Aubourg}},\ }\emph {\bibinfo {title} {{Etudes exp{\'{e}}rimentales de la
  turbulence d'ondes \`{a} la surface d'un fluide. La th\'{e}orie de la
  Turbulence Faible \`{a} l'\'{e}preuve de la r\'{e}alit\'{e} pour les ondes de
  capillarit\'{e} et gravit\'{e}}}},\ \href@noop {} {Ph.D. thesis},\ \bibinfo
  {school} {Universit\'{e} Grenoble Alpes} (\bibinfo {year} {2016})\BibitemShut
  {NoStop}%
\bibitem [{\citenamefont {Scott}(2014)}]{Scott}%
  \BibitemOpen
  \bibfield  {author} {\bibinfo {author} {\bibfnamefont {J.F.}\ \bibnamefont
  {Scott}},\ }\bibfield  {title} {\enquote {\bibinfo {title} {{Wave turbulence
  in a rotating channel}},}\ }\href@noop {} {\bibfield  {journal} {\bibinfo
  {journal} {J. Fluid Mech.}\ }\textbf {\bibinfo {volume} {741}},\ \bibinfo
  {pages} {316--349} (\bibinfo {year} {2014})}\BibitemShut {NoStop}%
\end{thebibliography}%
%---------------------------------------------------------------------------------------
\end{document}